\documentclass[lettersize,journal]{IEEEtran}
\usepackage{amssymb}  
\usepackage{url}
\usepackage{subcaption}
\usepackage{multirow}
\usepackage{booktabs}
\usepackage{graphicx}
\usepackage{algorithm}
\usepackage{algorithmic}
\usepackage{threeparttable}
\usepackage{float}
\usepackage{pifont}
\usepackage{makecell}
\usepackage{enumitem}
\usepackage{wrapfig}
\usepackage{xcolor}
\usepackage{hyperref}

\usepackage{xurl}

\usepackage{xspace}

\newcommand{\Name}{\texttt{TESLA}\xspace}

\begin{document}

\title{Capacitive Touchscreens at Risk: A Practical Side-Channel Attack on Smartphones via Electromagnetic Emanations}

\author{Yukun Cheng, Changhai Ou, Shiyu Zhu, Jinyuan Zhang, Zhenfang Qiu, Xingshuo Han,

Tianwei Zhang, \IEEEmembership {Member, IEEE}, Yuan Li and Shihui Zheng
\thanks{This work was supported by the National Key R\&D Program of China under Grant 2022YFB3103800, in part by the National Cryptologic Science Fund of China under Grant 2025NCSF02056, in part by the Natural Science Foundation of Wuhan under Grant 2024040801020235, in part by the Fundamental Research Funds for the Central Universities under Grant 2042025kf0051 and in part by Henan Key Laboratory of Network Cryptography Technology under Grant LNCT2025006. Corresponding author: Changhai Ou.}

\thanks{Yukun Cheng, Changhai Ou, Shiyu Zhu, Jinyuan Zhang and Zhenfang Qiu are with the School of Cyber Science \& Engineering, Wuhan University, Wuhan 430072, Hubei, China and Henan Key Laboratory of Network Cryptography Technology, Zhengzhou 450000, Henan, China (e-mail: kuin33@whu.edu.cn; ouchanghai@whu.edu.cn; zhushiyu@whu.edu.cn; 2023282210268@whu.edu.cn; 2018302180151@whu.edu.cn).

Xingshuo Han is with the College of Computer Science and Technology, Nanjing University of Aeronautics and Astronautics, Nanjing 211106, Jiangsu, China (e-mail: xingshuo.han@nuaa.edu.cn).

Tianwei Zhang is with the School of Computer Science and Engineering, Nanyang Technological University, Singapore 639798 (e-mail: tianwei.zhang@ntu.edu.sg).

Yuan Li is with the College of Computer, National University of Defense Technology, Changsha 410003, Hunan, China (e-mail: liyuan.margret@hotmail.com).

Shihui Zheng is with the School of Cyberspace Security, Beijing University of Posts and Telecommunications, Beijing 100876, China (e-mail: shihuizh@bupt.edu.cn).}}

\markboth{IEEE Transactions on Mobile Computing}%
{Cheng \MakeLowercase{\textit{et al.}}: Capacitive Touchscreens at Risk: A Practical Side-Channel Attack on Smartphones via Electromagnetic Emanations}

\IEEEpubid{0000--0000/00\$00.00~\copyright~2021 IEEE}

\maketitle

\begin{abstract}
Capacitive touchscreens in modern smartphones introduce severe side-channel vulnerabilities. However, existing attacks often require restrictive conditions or invasive measurements. This paper presents \Name, a novel, contactless electromagnetic (EM) side-channel attack that exploits inherent EM emanations during touchscreen scanning. We demonstrate that these emanations encode the spatiotemporal evolution of touch interactions, forming a unified leakage basis. By secretly placing an EM probe near the victim's device, \Name enables attackers to extract highly sensitive information, including screen-unlocking PIN codes, keyboard inputs, interacting application categories, and continuous handwriting trajectories. Compared to existing attacks, \Name offers a broader range of attack targets, more efficient sample acquisition, and operations in practical attack scenarios. Extensive evaluations on popular commercial smartphones, specifically the iPhone X, Xiaomi 10 Pro, Samsung S10, and Huawei Mate 30 Pro, validate the effectiveness of \Name. It achieves remarkable inference accuracy in diverse settings such as private meeting rooms and public libraries, with success rates of 99.3\% for PIN code recognition, 97.6\% for keyboard input reconstruction, and 95.0\% for application inference,  respectively. Simultaneously, it attains a 76.8\% character recognition accuracy and a high geometric similarity (Jaccard index of 0.74) for 2D handwriting trajectory reconstruction\footnote{Demo videos can be found on our anonymous website: \url{https://zenodo.org/records/15212428}.}.

\end{abstract}

\begin{IEEEkeywords}
Electromagnetic side channels, Capacitive touchscreen, Privacy leakage, Mobile security
\end{IEEEkeywords}

\begin{figure}[!t]
  \centering
  \includegraphics[width=0.95\linewidth]{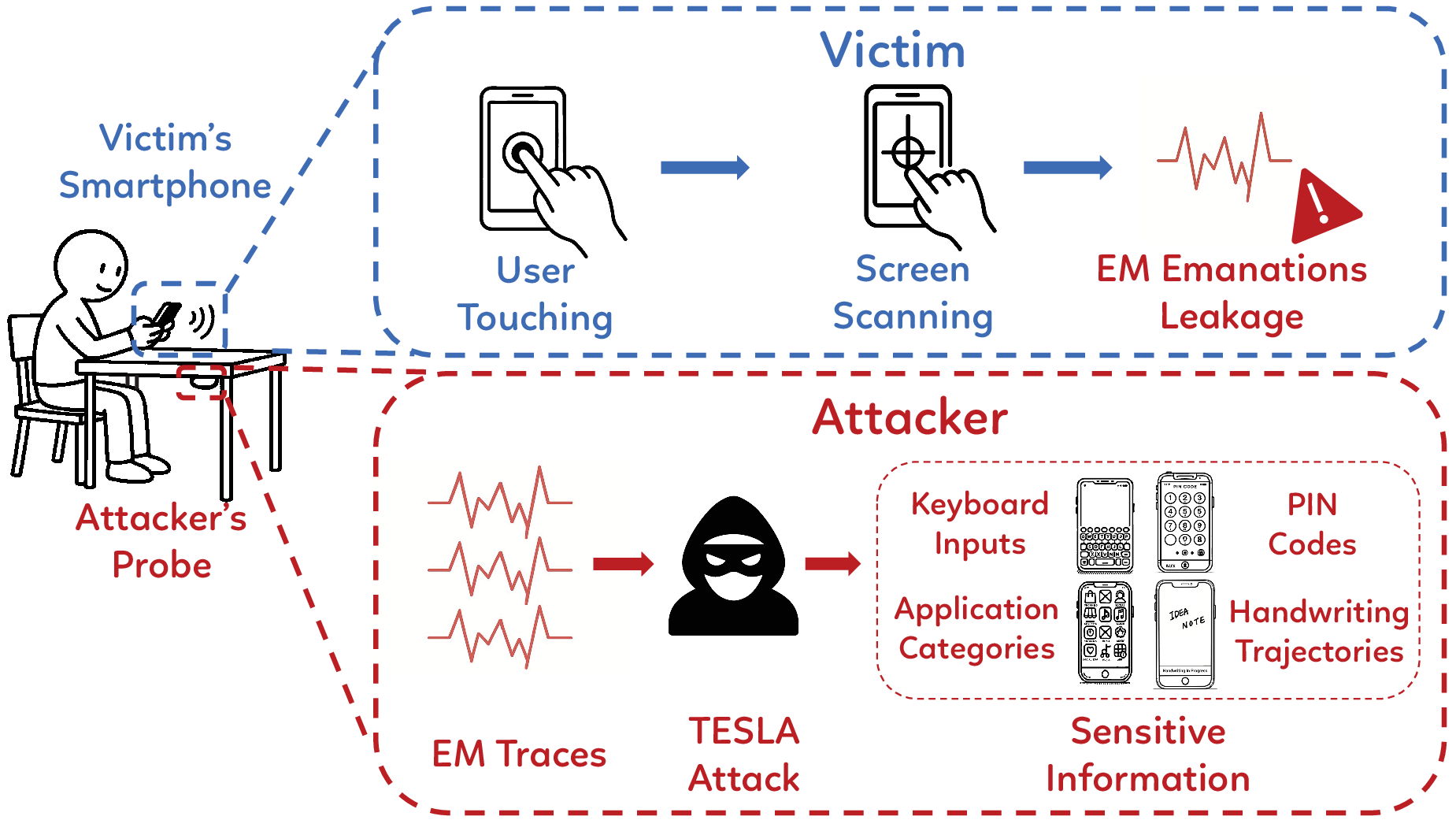}
  \caption{\Name exploits the EM emanations generated during the user touch interactions to recover sensitive user information, including keyboard inputs, PIN codes, application categories, and handwriting trajectories.}\label{short overview}
\end{figure}

\begin{table*}[!tbp]
\caption{Comparison with representative side-channel attack methods against smartphones.}
\resizebox{\linewidth}{!}{\begin{tabular}{c|l|l|c|c|c}
\Xhline{1.5pt}
\textbf{Attack Method}                              & \multicolumn{1}{c|}{\textbf{Side Channel and Information}}                                                                & \multicolumn{1}{c|}{\textbf{Attack Targets}}                                                                               & \textbf{\begin{tabular}[c]{@{}c@{}}Number of  Samples\\ per Minute Signal\end{tabular}} & \textbf{\begin{tabular}[c]{@{}c@{}}Non-Intrusive \\ Measurement\end{tabular}} & \textbf{Condition-Agnostic}                    \\ \Xhline{1.5pt}
FPLogger \cite{NiZZ23}             & EM emanations during the fingerprint sensor scanning   & \multicolumn{1}{l|}{1. Fingerprints}                                                                                      & 30                                    & \textbf{$\times$}                  & \textbf{$\times$} (Wireless Charging)     \\ \hline
Periscope \cite{jin2021periscope}  & EM emanation with finger movements over the touchscreen & \multicolumn{1}{l|}{1. PIN Codes}                                                                                         & 75                                    & \textbf{$\checkmark$}              & \textbf{$\checkmark$} \\ \hline
Charger-Surfing \cite{Cronin0YW21} & Power consumption related to the activities on the touchscreen & \multicolumn{1}{l|}{1. PIN Codes}                                                                                         & 120                                   & \textbf{$\times$}                  & \textbf{$\times$} (USB Charging)          \\ \hline
Yang et al. \cite{YangGZFB17}      & Power consumption related to the Webpage loading  & \multicolumn{1}{l|}{1. Loading Webpages}                                                                                  & 10                                    & \textbf{$\times$}                  & \textbf{$\times$} (USB Charging)          \\ \hline
WISERS \cite{NiZZLYWXLZ23}         & \begin{tabular}[c]{@{}l@{}}Coil whine and magnetic field perturbation of user \\ interactions on the charging smartphone\end{tabular}  & \begin{tabular}[c]{@{}l@{}}1. User Interactions \\ 2. PIN Codes \\ 3. Keyboard Inputs\end{tabular} & 600                                   & \textbf{$\checkmark$}              & \textbf{$\times$} (Wireless Charging)     \\ \hline
Radsee \cite{ZhangWG0Z25}      & Radio signals during the victim’s handwriting  & \multicolumn{1}{l|}{1. Handwriting Trajectories}                                                                                  & 85                                    & \textbf{$\checkmark$}                  & \textbf{$\times$} (Specialized Radar Hardware)          \\ \hline
\textbf{\Name}                                      & EM emanations during user-touch interactions & \begin{tabular}[c]{@{}l@{}}1. PIN Codes \\ 2. Keyboard Inputs \\ 3. Application Categories\end{tabular} & 7200 or 10800                         & \textbf{$\checkmark$}              &\textbf{$\checkmark$} \\ \Xhline{1.5pt}
\end{tabular}}
\label{tab: compare}
\end{table*}

\section{Introduction}
\IEEEpubidadjcol
\IEEEPARstart{C}APACITIVE screens are widely used in commercial smartphones from various mainstream manufacturers, such as Apple, Samsung, Huawei, and Xiaomi. Compared to traditional resistive screens, capacitive screens offer the advantages of supporting multi-touch functionality, higher durability, and more responsive touch feedback.
However, recent studies have shown that capacitive touchscreens inadvertently introduce serious privacy vulnerabilities due to various side channels. An adversary can exploit such information leakages during user-mobile interactions to recover sensitive user data, e.g., fingerprints \cite{NiZZ23}, screen-unlocking PIN codes \cite{Cronin0YW21, Zhuoran2021Screen, jin2021periscope}, detailed interaction patterns \cite{NiZZLYWXLZ23}, and handwriting trajectories \cite{ZhangWG0Z25, WeiZ15}. For example, the power consumption trace of a charging smartphone reveals dynamic content on its screen, which can be used to determine the location of animations triggered by button presses \cite{Cronin0YW21}. Similarly, an attacker can use an antenna and a software-defined radio (SDR) to intercept the electromagnetic (EM) signals emitted by the device to the screen and reconstruct them into grayscale images \cite{Zhuoran2021Screen}.

Although existing work has established the technical feasibility of side-channel attacks against smartphones, several limitations hinder their practical effectiveness. 
(i) \textit{Limited attack capability}. Current methods mainly target specific types of sensitive information \cite{jin2021periscope, Cronin0YW21} rather than comprehensive user activities, reducing the generality of the attack. (ii) \textit{Excessive temporal requirements}. The requirement of prolonged signal capture durations imposes heavy sampling overheads for attackers \cite{NiZZ23, YangGZFB17}. (iii) \textit{Intrusive measurement}. Several existing highly-accurate attacks required the modifications of physical devices for measurement \cite{Cronin0YW21, NiZZ23}, significantly increasing the complexity of attack deployment. (iv) \textit{Laboratory-condition dependency}. Reliance on specific device states \cite{NiZZLYWXLZ23, YangGZFB17} or custom-designed measurement hardware \cite{ZhangWG0Z25, WeiZ15} reduces the authenticity of the attack model.

To overcome these limitations, this paper proposes a practical and novel contactless \textbf{T}ouchscreen \textbf{E}lectromagnetic \textbf{S}ide-channel \textbf{L}eakage \textbf{A}ttack (\Name). 
We bring to light an unexplored security vulnerability: as shown in Fig. \ref{short overview}, an adversary can simply place a probe near the victim's smartphone to capture EM emanations generated during the user-touch interactions, thereby recovering sensitive user information effectively. The root cause of this vulnerability lies in the human coupling principle of the capacitive touchscreen and the spatial dependencies of the screen scanning driving method. Through extensive analysis, we discover that these EM emanations contain sufficient information to reconstruct highly sensitive user data, including screen-unlocking PIN codes, keyboard inputs, application categories, and handwriting trajectories. Notably, while everyday typing relies heavily on virtual keyboards, continuous handwriting on touchscreens remains a critical modality for high-security tasks, such as verifying electronic signatures in financial banking applications and signing legally binding PDF contracts.

\Name operates via a carefully-designed four-phase pipeline. First, in the user interaction classification phase, the attacker identifies the user behavior based on the collected EM signals to determine the attack target. In the subsequent signal interception phase, a signal interception algorithm detects the start and end positions of the valid signal and intercepts it. Next, the signal normalization phase reshapes and normalizes the signal to ensure optimal input for sensitive information extraction. Finally, the normalized signals are fed into the pre-trained deep learning model within the sensitive information extraction phase, which leverages its learned feature extraction capabilities to classify and reconstruct various types of sensitive information, including screen-unlock PINs, keyboard inputs, interacting application categories, and handwriting trajectories.

Compared to existing methods, \Name offers the following advantages, as detailed in Table~\ref{tab: compare}. (i)
\Name accurately identifies touch positions from EM traces during single touchscreen events, thereby enabling the targeting of a broader range of sensitive user information, such as screen-unlocking PIN codes, keyboard inputs, interacting application categories, and handwriting trajectories. This represents an improvement over existing attacks, which have typically focused on a single target (i.e., PIN codes) due to the complexity of information recovery methods \cite{Cronin0YW21, jin2021periscope}. (ii) Unlike existing attacks \cite{NiZZ23, Cronin0YW21}, which were limited to obtaining only one sample from a single long-duration sampling, e.g., 0.5 $\sim$ 10 seconds, \Name collects more samples within the same time period, e.g., 7200 or 10800 samples per minute of signal. This approach substantially enlarges the attacker's capabilities, as it enables the acquisition of more samples within the same time frame. (iii) \Name does not require modifications to existing physical devices, e.g., compromising USB cables \cite{Cronin0YW21} or power stations to monitor power traces \cite{YangGZFB17}. Instead, an attacker can execute the attack by simply placing a measurement probe in close proximity, e.g., 5 cm, to the target device. (iv) Unlike prior works that relied on specific conditions, e.g., requiring the victim's smartphone to be placed on a wireless charger \cite{NiZZLYWXLZ23},  \Name eliminates the need for restrictive preconditions. These advantages make our attack more feasible and practical in real-world scenarios.

To evaluate the practicality of \Name, we conduct extensive evaluations on several off-the-shelf smartphones from four manufacturers, i.e.,  iPhone X, Xiaomi 10 Pro, Samsung S10, and Huawei Mate 30 Pro. We perform practical end-to-end attacks to extract sensitive information from both discrete sequences of user interactions and continuous fine-grained handwriting. Experimental results demonstrate that \Name achieves remarkable performance. For discrete inputs, it successfully reveals sensitive information with a minimum of 95.33\% success rate in recovering complete user input sequences within five attempts (top-5). Furthermore, for continuous handwriting trajectory recovery, it attains a 77\% character recognition accuracy and a high geometric similarity with a Jaccard index of 0.74.

We summarize the contributions as follows:

\begin{itemize}[leftmargin=*,noitemsep,topsep=0pt,parsep=0pt,partopsep=0pt]
\item \textbf{A new capacitive screen security vulnerability.} 
We reveal an alarming security vulnerability in which sensitive user information may be inadvertently leaked through EM emanations from smartphone capacitive screens during touch interactions.

\item \textbf{A unified side-channel attack framework.} 
Based on the vulnerability, we propose a comprehensive side-channel attack framework, \Name, which is capable of extracting a wide spectrum of sensitive user information, encompassing both discrete inputs (e.g., screen-unlocking PIN codes, keyboard inputs, application categories) and continuous fine-grained handwriting trajectories, in a contactless manner.

\item \textbf{Physical world evaluations.} 
\Name is evaluated on four commodity smartphones from different manufacturers. Experimental results on both public and private environments highlight its effectiveness and adaptability in real-world scenarios. Moreover, our evaluations demonstrate that \Name is robust against a variety of practical impact factors.
\end{itemize}

This paper is an extended version of our previous work \cite{dac26} accepted in Design Automation Conference 2026. While the preliminary study focused solely on a specific attack vector targeting continuous handwriting trajectories, this manuscript fundamentally elevates the core scientific logic to propose a generalized attack framework. Specifically, we unify the underlying physical leakage mechanisms to decode both discrete inputs (e.g., PIN codes, keyboards) and continuous spatiotemporal inputs within a single, cohesive framework. To support this generalized architecture, we systematically designed targeted feature extraction methods varying significantly by fault models, integrating specialized Multi-Layer Perceptrons (MLP) and Convolutional Neural Networks (CNN) alongside the original sequence-based models. Furthermore, this manuscript introduces comprehensive end-to-end evaluations to assess the realistic cascading success rates of recovering complete multi-step input sequences under diverse environmental constraints.



\section{Preliminaries}
\subsection{Capacitive Touchscreen}


Capacitive and resistive touchscreens are the two most common technologies used in modern mobile devices. Resistive touchscreens detect touch through pressure, causing two Indium Tin Oxide (ITO) layers to make contact and change their resistance \cite{bhalla2010comparative}. In contrast, capacitive touchscreens detect touch by measuring changes in capacitance when fingers alter the electrostatic field \cite{kwon2018capacitive}. Nowadays, capacitive touchscreens have become dominant in smartphones and other mobile devices. This is due to their advantages over resistive screens, such as better durability, higher sensitivity, and the ability to support multi-touch capabilities \cite{Nam2021Review}.

\begin{figure}[t]
  \centering
  \includegraphics[width=0.95\linewidth]{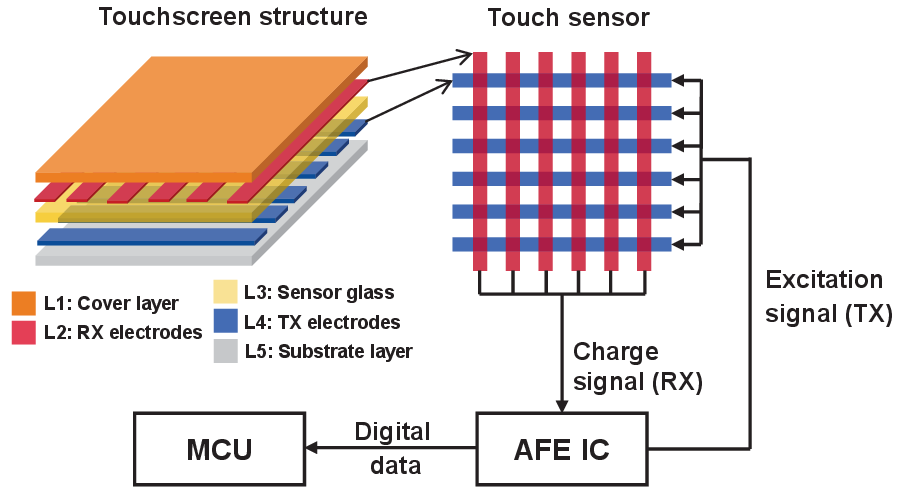}
  \caption{System architecture of capacitive touchscreen.}\label{screen}
  \vspace{-10pt}
\end{figure}

 Fig. \ref{screen} shows a typical architecture of a capacitive touch system \cite{jiang2023marionette}. The touchscreens usually contain five layers. Among these, the touch sensor is a vertical capacitor, composed of two ITO layers: the grid of transmitting (TX) and receiving (RX) electrodes. The touch sensor interacts with the analog front-end integrated circuit (AFE IC), which sends an excitation signal to the TX electrodes and processes the signals from the RX electrodes. The processed signal will be sent to the microcontroller unit (MCU) for further data processing and touch event detection.

 The operation of capacitive touchscreens is mainly based on the principle of mutual capacitance \cite{wang2022ghosttouch}. When the AFE IC sends the excitation signal to the TX electrodes, it creates an electrostatic field that interacts with the air-gapped RX electrodes and forms a mutual capacitance. When a finger approaches or touches the screen, it disturbs the electric field between the TX and RX electrodes, causing a change in the mutual capacitance. This change occurs because the finger acts as a conductor, drawing some of the charge from the electrodes. The change in capacitance affects the charge signal detected by the AFE IC, allowing the system to identify the touch point at the intersection of the TX and RX electrodes. 


\subsection{Human Coupling Effect}
\label{sec: Human Coupling Effect}
\begin{figure}[t]
\centering
\subfloat[]{\includegraphics[width=.45\linewidth]{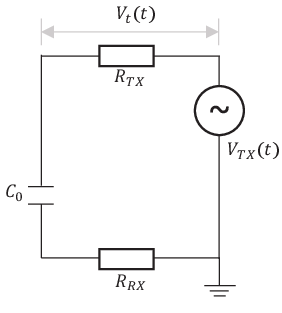}\label{touching}}%
\subfloat[]{\includegraphics[width=.45\linewidth]{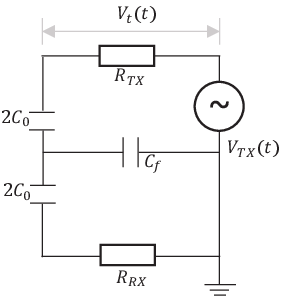}\label{no_touching}}
\caption{Illustration of human coupling effect in touchscreen. (a) Without touching. (b) With touching.}
\label{finger_coupling}
\vspace{-10pt}
\end{figure}

Based on the literature \cite{NiZZ23, jin2021periscope},  the finger-coupling effects that occur during a touch event on a capacitive touchscreen can be illustrated as Fig. \ref{finger_coupling}. The main driving voltage for electromagnetic emission of a capacitive touchscreen $V_t(t)$ is significantly influenced by this effect. In Fig. \ref{touching}, when no touch is present, the capacitive touch system is modeled by a simple circuit with a capacitance $C_0$ between the TX and RX electrodes, and $V_t(t)$ can be expressed as: 
\begin{equation}
  V_t(t) = V_{TX}(t) \cdot \frac{R_{TX}}{R_{TX} + R_{RX} + \frac{1}{j 2 \pi f C_0}},
\end{equation}
where $V_{TX}(t)$, $R_{TX}$ and $R_{RX}$ respectively denote the excitation signal and the resistor at TX and RX electrodes, $f$ denotes the frequency of the current, and $j$ denotes the imaginary unit.

In Fig. \ref{no_touching}, when a finger touches the screen, it introduces additional capacitance $C_f$, which results in a coupling effect that alters the system's impedance. In this case, the driving voltage $V_t(t)$ can be expressed as:
\begin{equation}
  V_t(t) = V_{TX}(t) \cdot \frac{R_{TX}}{R_{TX} + \frac{1}{j 4 \pi f C_0} + \Delta Z_f(t)},
\end{equation}
where the introduced impedance $\Delta Z_f(t)$ is formed by the parallel and series combination of the finger's coupling capacitance $C_f$ and the intrinsic circuitry. Based on the equivalent circuit model for mutual capacitance sensing \cite{jin2021periscope}, it can be derived and expressed as: can be expressed as:
\begin{equation}
  \Delta Z_f(t) = \frac{1}{\left( \frac{1}{1/{j 2 \pi f \Delta C_f}} + \frac{1}{1/{j 4 \pi f C_0 + R_{RX}} }\right)}.
\end{equation}

As the finger approaches the screen, the capacitance $C_f$ increases, leading to a stronger coupling effect and a reduction in impedance $\Delta Z_f(t)$. This change results in an increase in the driving voltage $V_t(t)$, amplifying the touchscreen’s electromagnetic emissions. Thus, the emitted signal varies with the finger's distance, providing a potential side channel for detecting touch interactions.


\subsection{Screen Scanning Methods}
\label{sec:screen-scanning}

\begin{figure}[t]
  \centering
  \includegraphics[width=\linewidth]{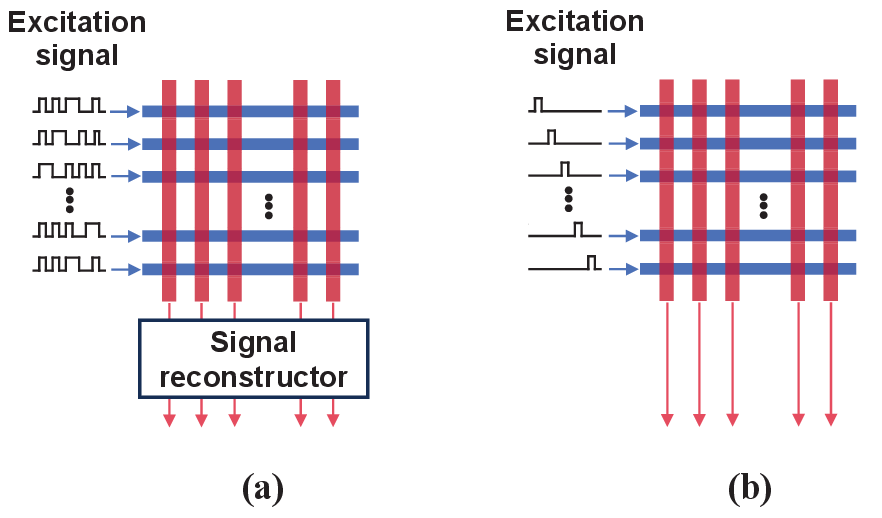}
  \caption{Driving methods in capacitive touchscreens. (a) Parallel driving method. (b)  Scan driving method.}
\label{driving_method}
  \vspace{-10pt}
\end{figure}

Parallel Driving Method (PDM) \cite{huang2019pdm} and Scan Driving Method (SDM) \cite{ko2010low} are two distinct screen scanning techniques employed in AFE ICs for detecting touch events and measuring changes in capacitance. The conceptual diagrams of SDM and PDM are summarized in Fig. \ref{driving_method}, as reviewed in \cite{Nam2021Review}.

PDM, as shown in Fig. \ref{driving_method}a, sends the excitation signals to all TX electrodes simultaneously. It has high SNR because the sensing time is multiplied by the number of TX electrodes, thus improving the quality of the detected signal. However, the increased sensing time comes at the cost of higher power consumption and greater circuit complexity. Since the excitation signals need to be applied simultaneously across all TX electrodes, orthogonal excitation patterns are used to prevent interference between signals.  Although PDM offers high SNR, it requires more power and a larger area due to the parallel operation of multiple excitation circuits.

As shown in Fig. \ref{driving_method}b, in SDM, the excitation signal is sequentially sent to the TX electrodes through a de-multiplexer (DEMUX), and the readout circuit measures the charge signals from the RX electrodes. The touch sensing time for each TX electrode is shared across all electrodes, resulting in time difference in the activation time of each electrode. SDM has the advantage of a simpler circuit design and a smaller footprint over PDM, making it more suitable for systems where minimizing area and cost is a priority. 

Most smartphones use SDM for their touchscreens due to its simple structure and short sensing time \cite{wang2022ghosttouch}. However, due to the feature of SDM, the excitation signal will have different time delays when scanning different positions on the screen. This feature provides the possibility for implementing side channel attacks against the capacitive touchscreen.

\subsection{Jaccard Index}
The Jaccard index was introduced by Swiss Jaccard to compare the distribution of flora across different geographic regions \cite{jaccard1912distribution}. It has since been widely adopted in various fields, including computer vision and data mining, for measuring the similarity between sample sets. Mathematically, the Jaccard index quantifies the similarity between two finite sample sets $A$ and $B$, and is defined as the ratio of the size of their intersection to the size of their union:

\begin{equation}
J(A, B) = \frac{|A \cap B|}{|A \cup B|} = \frac{|A \cap B|}{|A| + |B| - |A \cap B|}.
\end{equation}

The Jaccard index ranges from 0 to 1. A value of 1 indicates a perfect match, whereas a value of 0 indicates no overlap or similarity between the two sets. In the context of our attack, a higher Jaccard index reflects a more accurate recovery of the user's biometric handwriting features.

 \subsection{Side-channel Attacks on Smartphones}
 \label{sec:Related works}

\begin{figure*}[!tbp]
  \centering
  \subfloat[]{\includegraphics[width=.42\linewidth]{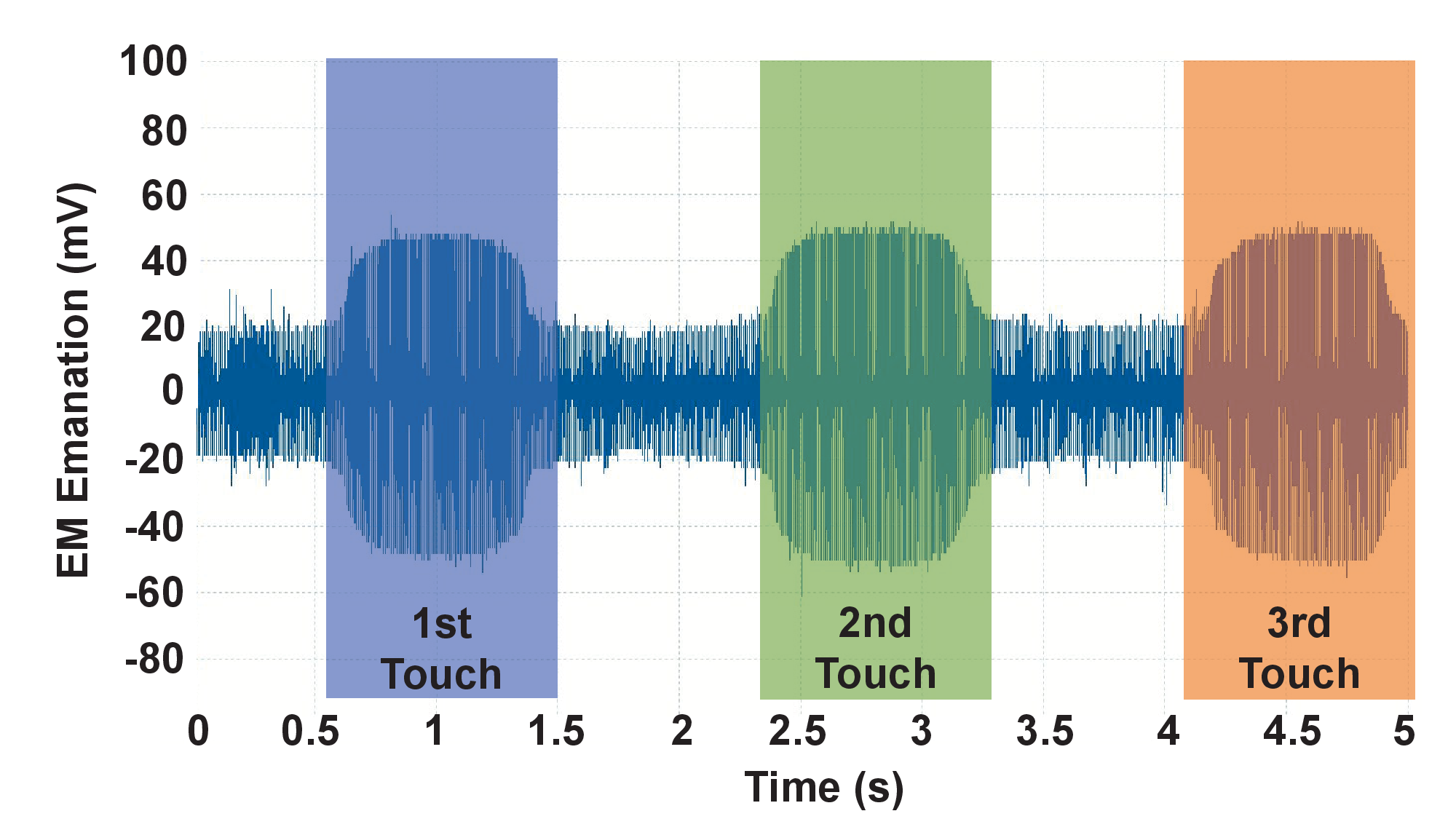}\label{three touch}}%
  \hspace{1cm}
  \subfloat[]{\includegraphics[width=.42\linewidth]{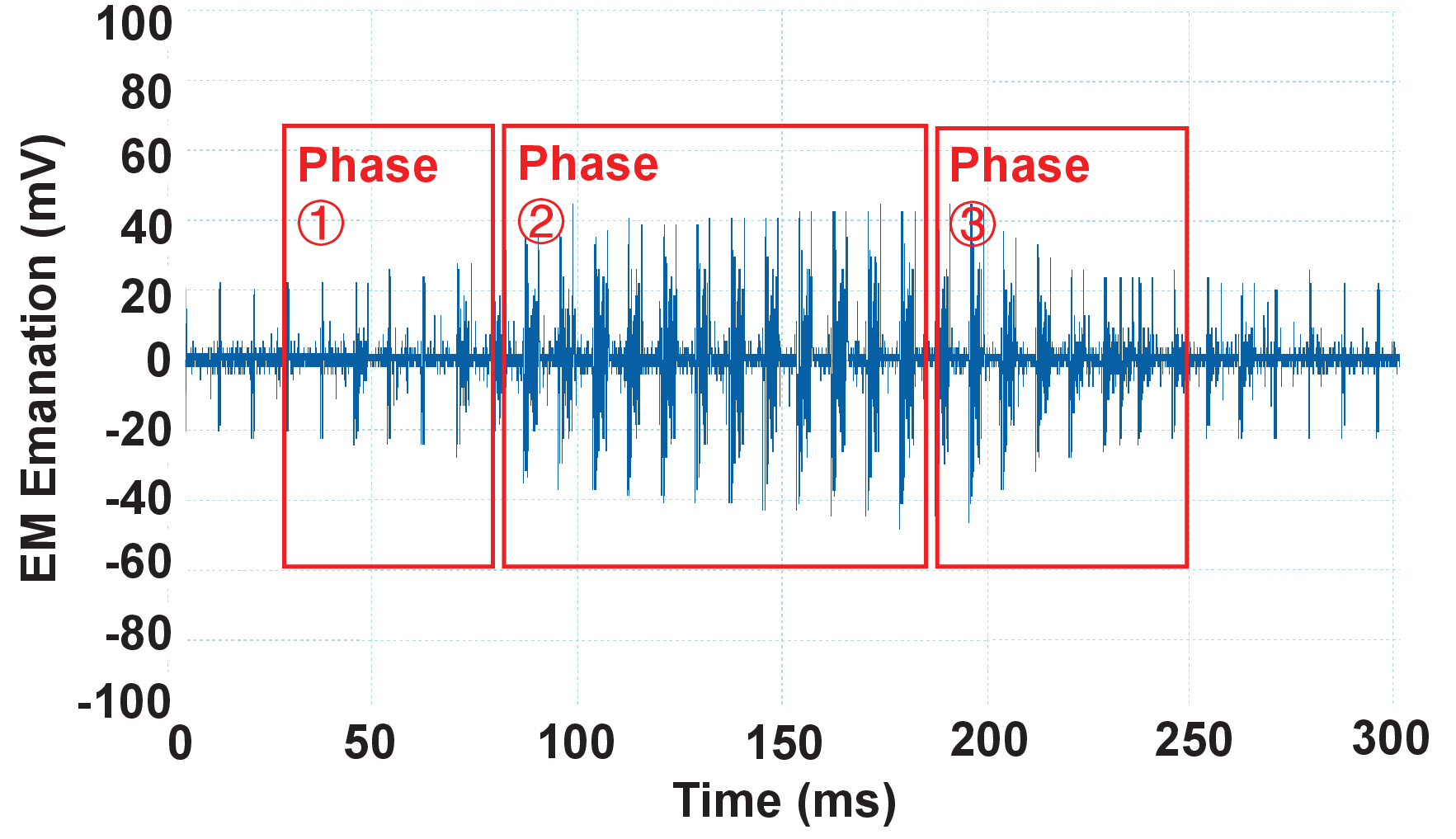}\label{one touch}}
  \caption{EM emanation measurements for touch interactions. (a) Three touch interactions. (b) Details of the first touch interaction with distinct phases corresponding to different user actions.}
\end{figure*}

\begin{figure}[t]
  \centering
  \includegraphics[width=\linewidth]{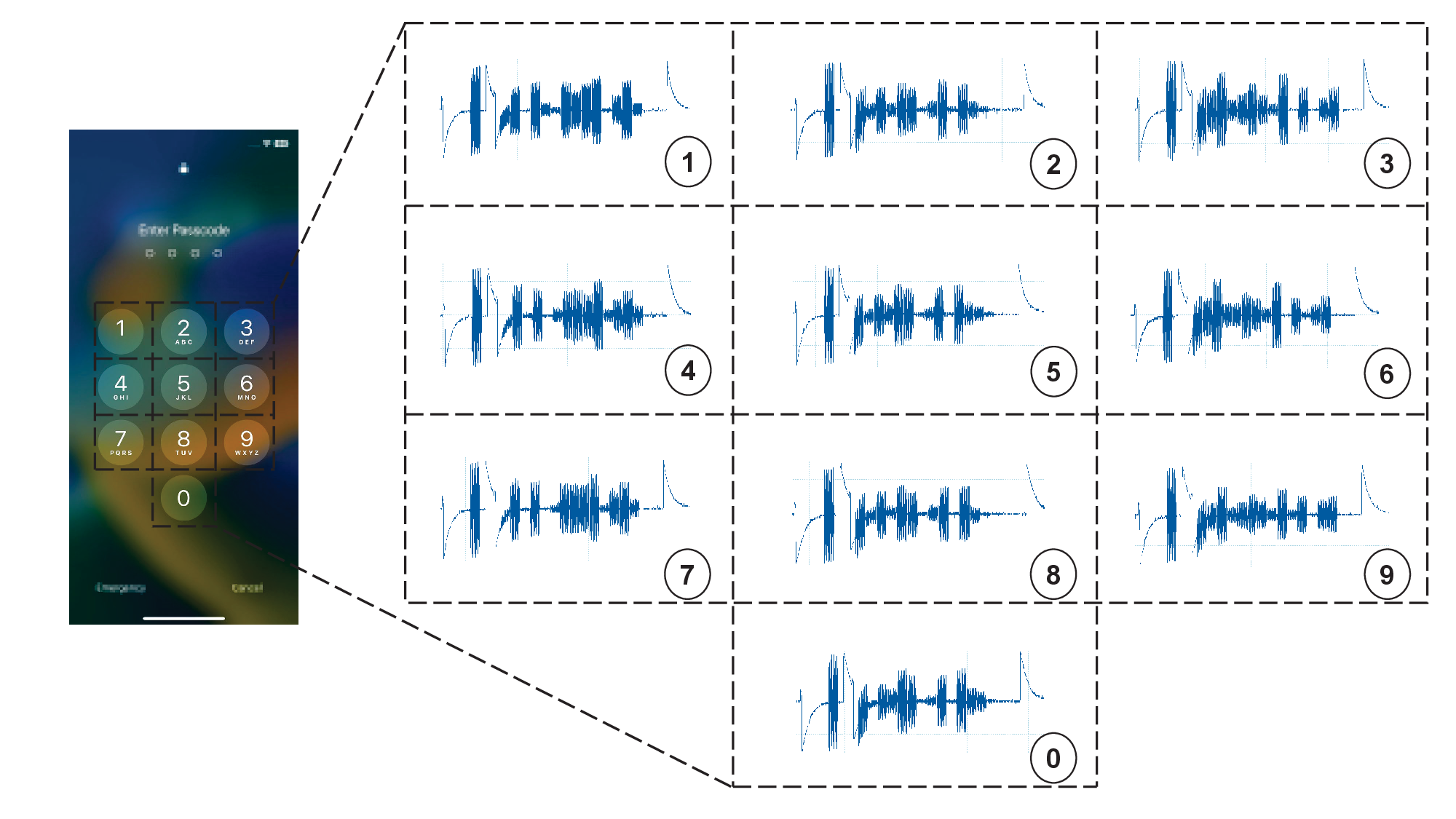}
  \caption{EM emanation measurements for different touch segments. The differences between horizontal segments are more pronounced than those among vertical segments.}\label{position}
  \vspace{-10pt}
\end{figure}

Existing side-channel attacks on smartphones can be broadly classified into three categories: sensor-based, power-based, and EM-based attacks.

\noindent\textbf{Sensor-based attacks.} 
Sensor-based side channels pose security threats by leveraging smartphone sensors, such as accelerometers or gyroscopes, to infer user interactions or compromise sensitive inputs like PINs \cite{Hussain2016Therise} or handwriting trajectories \cite{FarrukhYXYWC21}. Cai et al. \cite{CaiC11} first introduced the concept that distinct vibration patterns generated by tapping different screen locations can be used to infer keystrokes through motion data analysis, utilizing accelerometer and gyroscope sensors. This pioneering work spurred extensive research into side channels in smartphones, with subsequent studies exploring various built-in sensors, including accelerometers \cite{AvivSBS12, OwusuHDPZ12}, gyroscopes \cite{MichalevskyBN14, WangLC15, TaktakTK17}, motion sensors \cite{hsu2018indoor, Mehrnezhad2015TouchSignatures}, ambient-light sensors \cite{spreitzer2014pin}, and geomagnetic rotation vector sensors \cite{oberhuber2025power} as potential leakage sources. Sensor-based side-channel analysis typically requires a backdoor or hacking program to gain partial access to the victim's device data. Unlike existing approaches, \Name does not assume that the attacker has control over the victim's phone, making the attack more realistic. Furthermore, Ni et al. \cite{NiZZLYWXLZ23} innovatively proposed that attackers could use the magnetic sensor and microphone of another phone to infer keystrokes without requiring malicious software installation on the victim's phone. However, unlike this work, \Name does not rely on special conditions, e.g., the victim using a wireless charger, making it more broadly applicable.

\noindent\textbf{Power consumption-based attacks.} 
Attackers exploit variations in a device's power consumption to infer sensitive information without prior infiltration of the target device. Early power analysis typically leverage a compromised power cable, e.g., USB cable \cite{Spreitzer2018Systematic,Spolaor2017No}. A notable example is demonstrated by Yang et al. \cite{YangGZFB17}, who implemented a power monitoring device on the USB charging circuit to capture power consumption traces during webpage loading processes, achieving webpage recognition accuracy of exceeding 90\%. Cronin et al. \cite{Cronin0YW21} introduced Charger-Surfing, a power line side-channel attack that infers touchscreen activities, such as button presses, from power traces during USB charging. Their approach achieved up to 98.7\% accuracy for single button presses and 95.1\% for 4-digit passcode inference. While recent advancements \cite{NiZZLYWXLZ23,NiLZZWXLZ23} have demonstrated contactless power analysis using specialized near-field magnetic probes, these techniques often require precise spatial alignment with the device's Power Management IC and remain highly sensitive to internal computational noise. In contrast, \Name does not require physical contact, nor does it rely on precise PMIC alignment, making it more covert and adaptable to dynamic real-world environments.


\noindent\textbf{EM-based attacks.}
This type of attacks involve the exploitation of unintended EM emanations emitted by electronic devices to infer sensitive information without physical access to the device. These attacks typically rely on specialized equipment, such as coils and probes, to capture electromagnetic radiation leakage. Researchers have demonstrated that these attack methods can be used to obtain various types of sensitive information, including screen-unlocking PIN codes \cite{jin2021periscope}, secret input data \cite{Zhuoran2021Screen, Zhan2022Graphics, Ni2023Expliting}, cryptographic keys \cite{Monjur2021Nonce, Daniel2016ECDSA}, application usage patterns \cite{Lojenaa2024Ensuring, Cheng2019MagAttack} and handwriting trajectories \cite{ZhangWG0Z25, WeiZ15}. For example, Jin et al. \cite{jin2021periscope} introduced Periscope, an attack that leverages human-coupled electromagnetic emanations from mobile device touchscreens to infer user keystrokes, achieving a recovery rate of up to 61.7\% for 4-digit PINs. Furthermore, Ni et al. \cite{NiZZ23} demonstrated the feasibility of recovering fingerprints from in-display fingerprint sensors by exploiting electromagnetic emanations emitted during the unlocking process, achieving a significant success rate in spoofing fingerprint authentication systems. However, even the most recent EM-based side-channel frameworks \cite{LongJ0A00F24} often focus on an isolated, specific attack target and sometimes require strong assumptions, such as the phone being placed on a wireless charger\cite{NiZZ23}. In contrast, \Name establishes a generalized and unified attack framework. It not only eliminates restrictive physical preconditions but also successfully decodes both discrete inputs and continuous spatiotemporal movements under a single cohesive leakage model, making it significantly more aligned with complex real-world scenarios.

\section{Touch Interaction Leakage Exploration}

To comprehend the newly explored EM emanation leakage, we conduct an empirical study with an unmodified iPhone X as the target device. A measurement probe is positioned at a vertical distance of 5 cm close to the rear surface of the iPhone to capture EM traces, while an oscilloscope is used to monitor and record the signals. Unlike idealized shielded laboratory setups, our experiments were purposefully conducted in a typical, daily-use meeting room to reflect realistic threat models.

\begin{figure*}[t]
  \centering
  \subfloat[Touch interaction \textit{with} animated visual feedback.]{\includegraphics[width=0.46\linewidth]{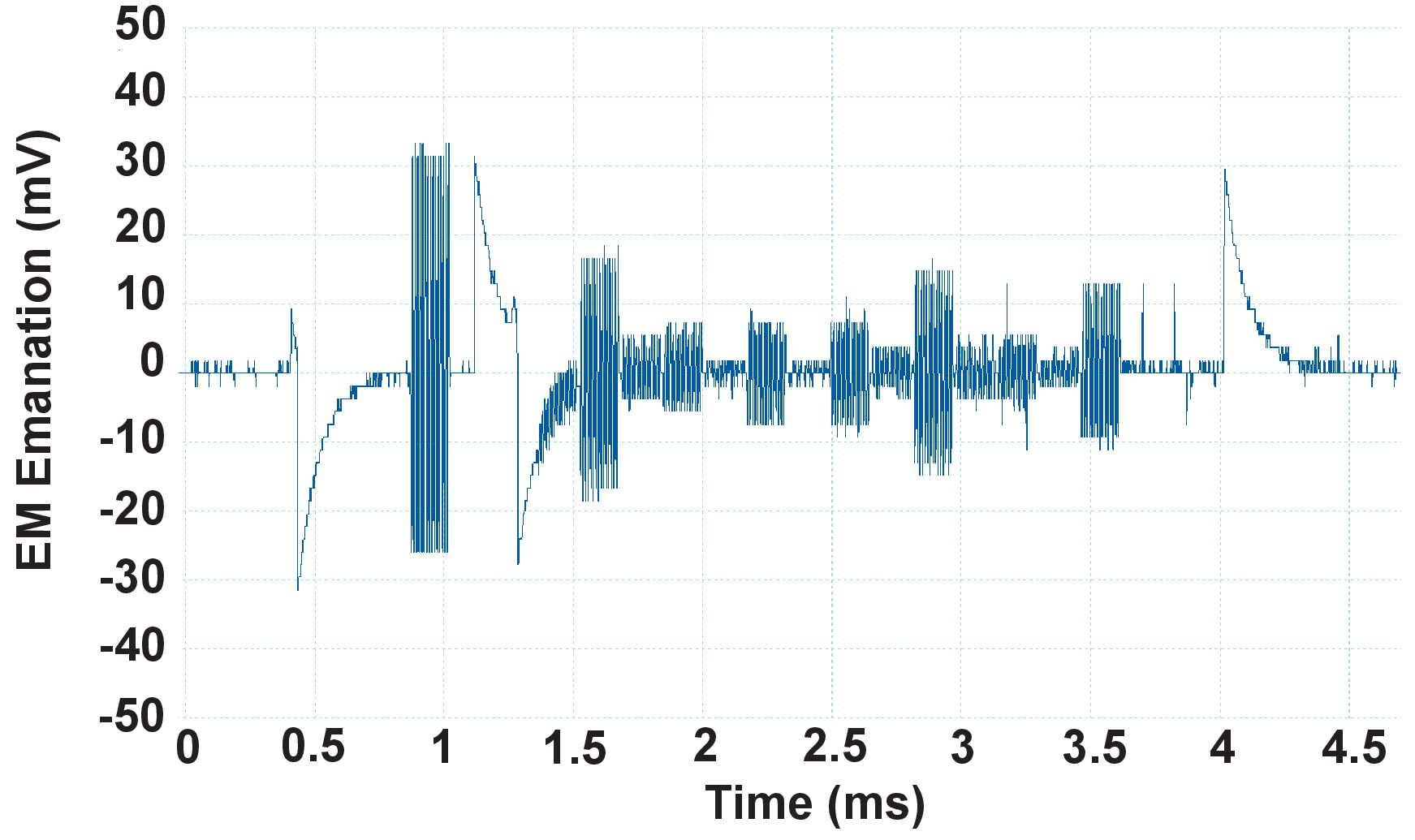}\label{white}}
  \hspace{1cm}
  \subfloat[Touch interaction \textit{without} animated visual feedback.]{\includegraphics[width=0.46\linewidth]{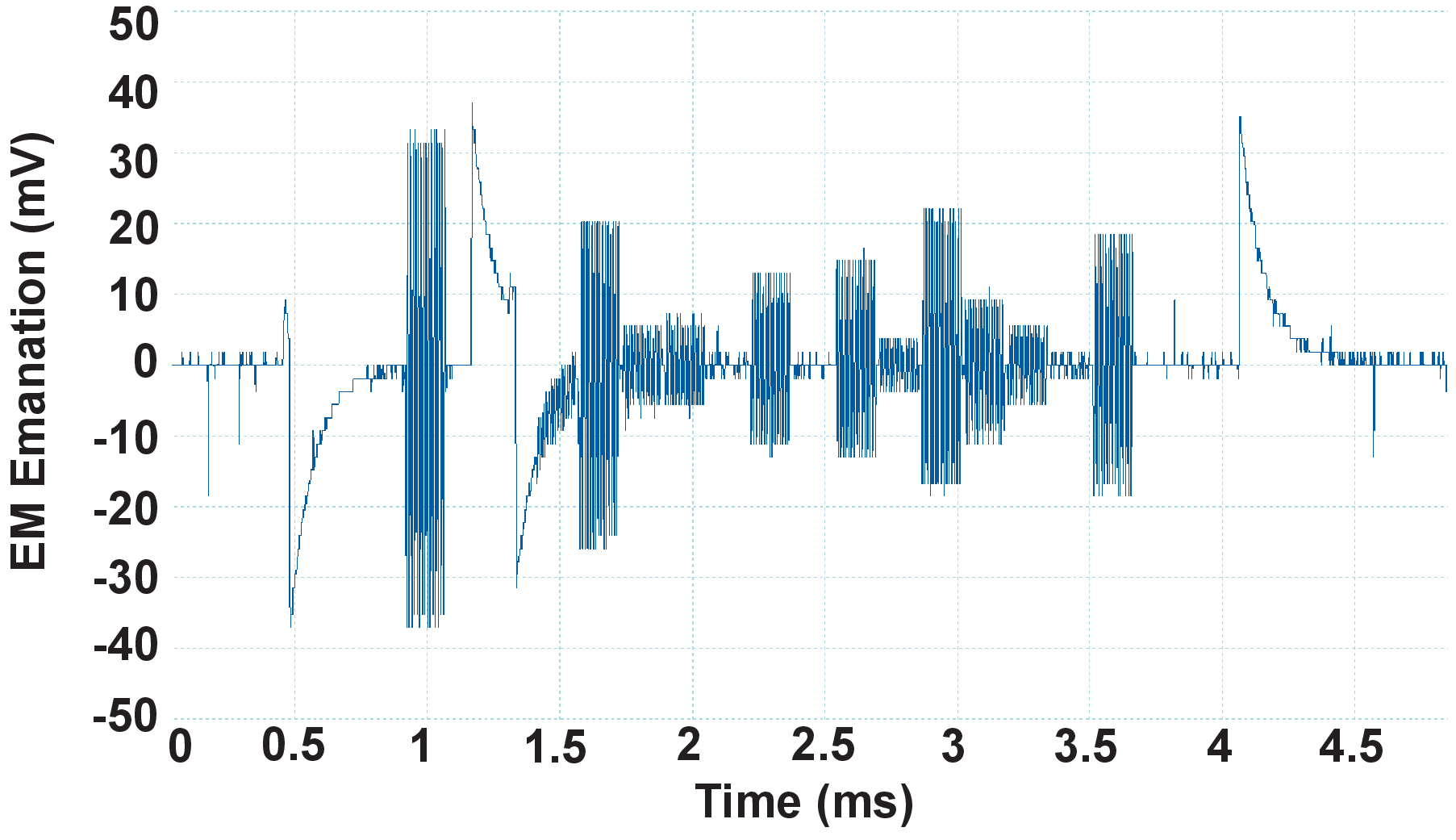}\label{black}}
  \caption{EM emanation measurements under different animated visual feedback conditions. The EM features of touch interaction are not affected by animated visual feedback conditions.}
  \label{fig:white-black}
\end{figure*}

\subsection{Touch Detection}
Our initial investigation focuses on detecting finger touch behavior through EM side-channel analysis. As illustrated in Fig. \ref{three touch}, when the smartphone screen remains untouched, fluctuations in the EM signal are minimal and primarily attributed to ambient noise. However, when a user's finger touches the screen, a distinct signal peak emerges, corresponding to the touch event.

Further, we analyze the details of the first touch interaction. As shown in Fig. \ref{one touch}, the touch interaction with the screen can be divided into three distinct phases. In phase \ding{172}, the user's finger gradually approaches the screen. It is evident that as the distance between the finger and screen decreases, the intensity of the EM emanations progressively increases. This observation aligns with the human coupling effect introduced in Section \ref{sec: Human Coupling Effect}. In phase \ding{173}, the user's finger touches the screen and remains for a period to complete the touch interaction. During this phase, the intensity of the EM emanations reaches its peak, and the signal waveform exhibits cyclical features. In phase \ding{174}, the user completes the touch interaction by gradually moving the finger away from the screen. It is observed that the signal strength diminishes as the distance between the finger and screen increases. These observations indicate that the captured EM emanations are strongly correlated with touchscreen interactions. Furthermore, the cyclical features observed in phase \ding{173} validate our investigation, and we will explore whether they lead to information leakage in Section \ref{sec: touch location}.



\subsection{Touch Location Identification}
\label{sec: touch location}
Our subsequent investigation aims to determine whether the cyclical features observed during touch interactions could cause information leakage. To this end, we divide the touchscreen into a 10-grid layout consisting of equal-area segments, which is based on the default keyboard layout corresponding to screen-unlocking PIN codes (shown in Fig. \ref{position}). We then conduct touch interactions at each segment and capture the corresponding EM traces. 

The captured EM traces for iPhone X are presented in Fig. \ref{position}, where a single feature cycle is highlighted to emphasize subtle variations in the signals. We observe that touches on different screen segments produce significantly distinct EM emanation patterns. Notably, the differences between horizontal segments (e.g., 1, 2, and 3) are more pronounced compared to the variations among vertical segments (e.g., 1, 4, and 7).  These results indicate that when the finger touches the screen, the EM emanations emitted by the phone will potentially leak information about the touch location.


\subsection{Leakage Mechanism Analysis}
Cronin et al.~\cite{Cronin0YW21} revealed that animated visual feedback induces power side-channel leakages of the touchscreen with similar cyclical features through position-dependent variations in animation rendering times, as empirically validated through charging power traces. In contrast, our investigation identifies a fundamentally different EM leakage phenomenon that operates independently of animated visual feedback systems. Fig. \ref{fig:white-black} compares touch interactions at the same screen location under two conditions: the normal state and a screen with black wallpaper where animation rendering is disabled. This comparison aims to verify the impact of animated visual feedback on EM leakage. Our observations confirm that the cyclical features remain consistent regardless of whether animated visual feedback is present or absent.

\begin{table*}[t]
\centering
\caption{The screen-related parameters of the smartphones and the frequency of the cyclical feature.}
\renewcommand{\arraystretch}{1}
\begin{tabular}{c|c|c|c|c|c}
\Xhline{1.5pt}
\textbf{Smartphone} & \textbf{Screen Material} & \textbf{Screen Vendor} & \textbf{\begin{tabular}[c]{@{}c@{}}Screen Refresh \\ Frequency (Hz)\end{tabular}} & \textbf{\begin{tabular}[c]{@{}c@{}}Touch Sampling \\ Frequency (Hz)\end{tabular}} & \textbf{\begin{tabular}[c]{@{}c@{}}Cyclical Feature \\ Frequency (Hz)\end{tabular}} \\ \Xhline{1.5pt}
iPhone 7            & LCD                & LG Display             & 60                                                                                & 60                                                                                & 60                                                                                  \\
iPhone X            & OLED               & Samsung             & 60                                                                                & 120                                                                                & 120                                                                                  \\
Xiaomi 10 Pro       & AMOLED                     & Samsung                & 90                                                                                & 180                                                                               & 180                                                                                 \\
Samsung S10         & AMOLED                   & Samsung                & 60                                                                                & 120                                                                               & 120                                                                                 \\
Huawei Mate30 Pro   & Dynamic AMOLED           & BOE                    & 60                                                                                & 120                                                                               & 120                                                                                 \\ \Xhline{1.5pt}

\end{tabular}
\label{table: frequency}
\end{table*}

To formally quantify the root cause of this vulnerability, we establish an analytical leakage model based on the touchscreen's hardware architecture. Modern smartphones predominantly employ the Scan Driving Method (SDM), where TX electrodes are activated sequentially within each sampling cycle. Assuming the touchscreen’s SDM architecture utilizes $N_{TX}$ vertical transmitting electrodes (columns) and $N_{RX}$ horizontal receiving electrodes (rows), let $T_{scan}$ denote the total time required to complete one full frame scan. Since the vertical TX electrodes are sequentially excited across the screen’s width, a touch event occurring at a spatial location with horizontal coordinate $x$—where $x \in [1, N_{TX}]$ represents the index of the active vertical TX column—will introduce a predictable temporal delay $\Delta t$ relative to the start of the scanning cycle. Based on the equivalent circuit model for mutual capacitance sensing \cite{ShanZZSWJ22}, this time-to-space mapping can be expressed as: 

\begin{equation}
    \Delta t(x) = \frac{x - 1}{N_{TX}} \cdot T_{scan} + \delta_{proc},
\end{equation}

\noindent where $\delta_{proc}$ represents a constant hardware processing offset. Consequently, the unique temporal shift $\Delta t(x)$ dictates the precise phase at which the maximum EM amplitude perturbation (driven by the impedance change $\Delta Z_f(t)$ in Eq. 3) occurs within the captured EM trace.

Crucially, this analytical model elucidates the fundamental hardware limitation observed in our empirical study in Section \ref{sec: touch location}: the diminished distinguishability between vertically adjacent regions (e.g., keys 1, 4, and 7) compared to horizontally adjacent ones (e.g., keys 1, 2, and 3). Horizontally separated keys lie on different vertical TX electrodes (i.e., distinct $x$ coordinates), resulting in pronounced and easily classifiable temporal shifts $\Delta t(x)$. Conversely, vertically aligned keys lie on the same vertical TX electrode. Because they share identical $x$ coordinates in the TX activation sequence, they are excited simultaneously. Their temporal delays are virtually identical, yielding highly similar EM emission envelopes. Distinguishing these vertical touches relies exclusively on much weaker secondary leakages along the horizontal RX lines, making them inherently more challenging for side-channel classifiers to separate.

This analytical model indicates that the temporal features of the EM leakage are intrinsically tied to the hardware scanning cycle. To further confirm that this leakage is indeed driven by the touchscreen's scanning hardware, we perform a frequency-domain parameter analysis. As shown in Table \ref{table: frequency}, according to the official technical documentation \cite{APPLE23, MI25, Sam21, Hua22}, the cyclical feature frequency is directly equal to the touch sampling frequency (associated with the screen's hardware response speed) and independent of the screen refresh frequency (associated with animation rendering). This consistency across multiple manufacturers pointing to the underlying scanning architecture as the primary source of leakage.

\begin{figure}[!tbp]
  \centering
  \subfloat[Attack in a library.]{\includegraphics[width=.46\linewidth]{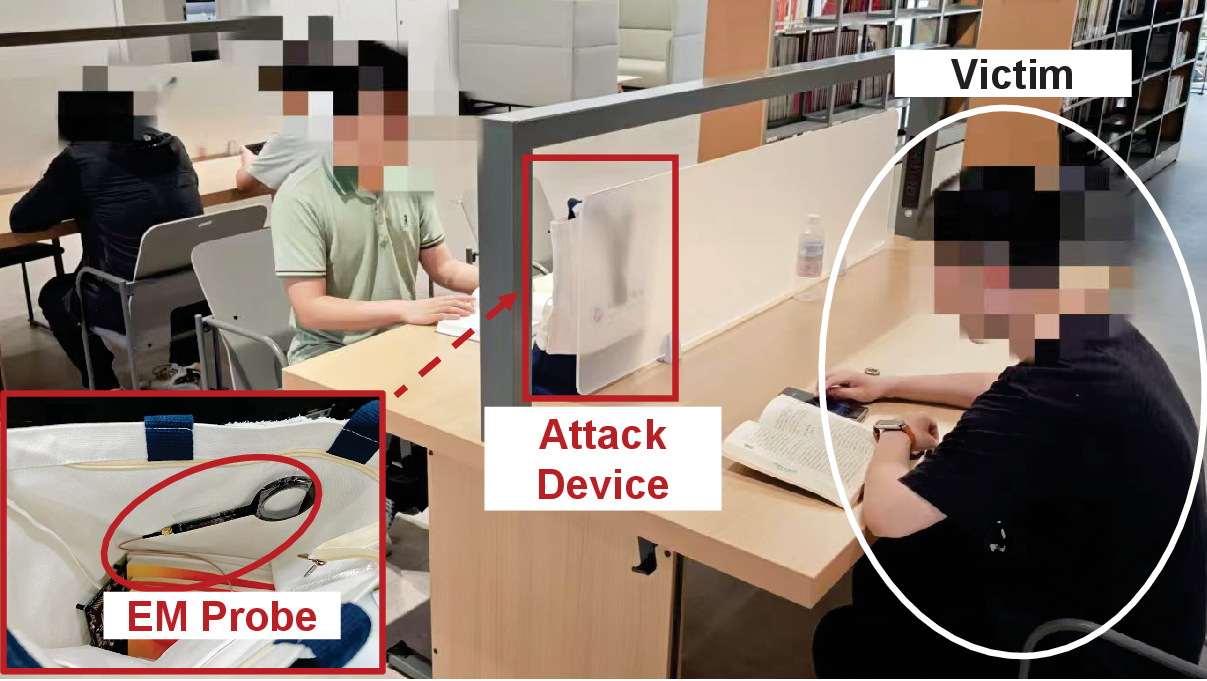}}%
  \subfloat[Attack in a meeting room.]{\includegraphics[width=.46\linewidth]{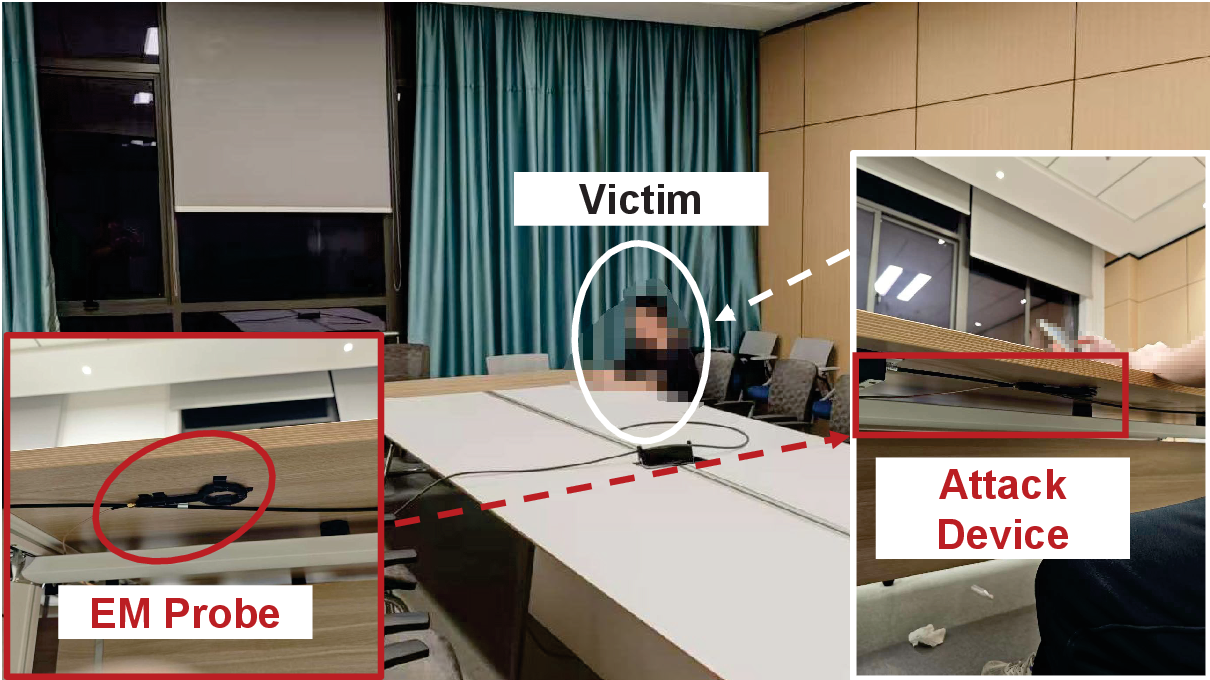}}
  \caption{Attack scenarios in the \textit{public} area and the \textit{private} area.}\label{environment}
\end{figure}

\begin{figure*}[!tbp]
  \centering
  \includegraphics[width=\linewidth]{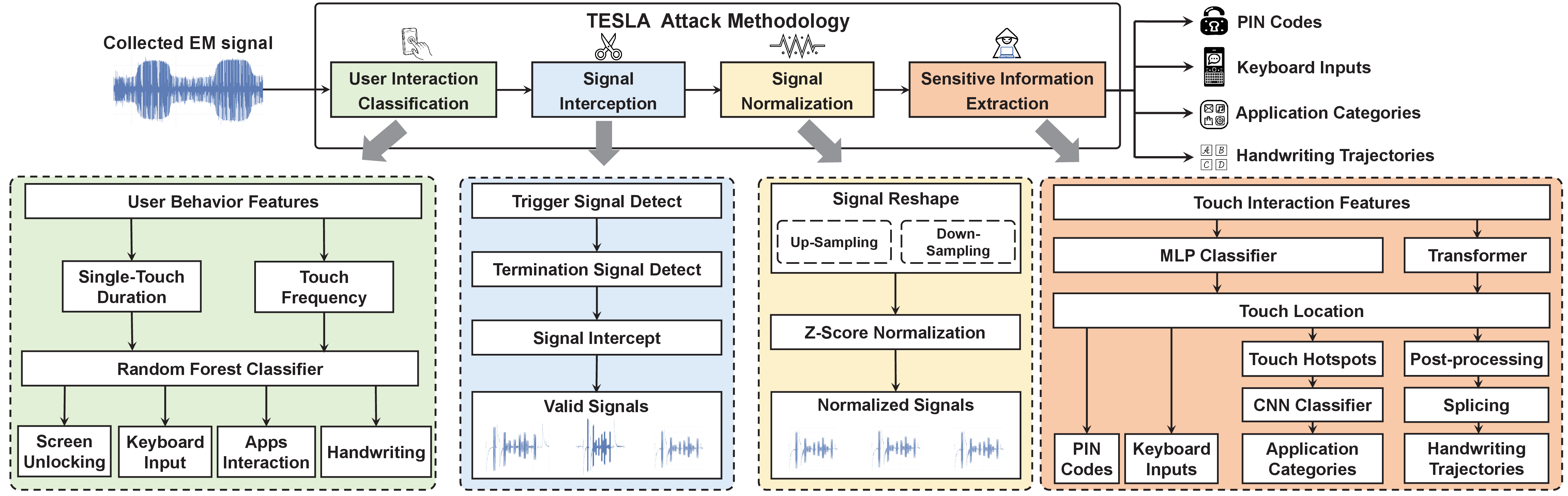}
  \vspace{-15pt}
  \caption{Overview of the \Name.}\label{Overview}
\end{figure*}

\section{Threat Model}
\label{sec: Threat Model}

\begin{figure}[!tbp]
  \centering
  \includegraphics[width=0.9\linewidth]{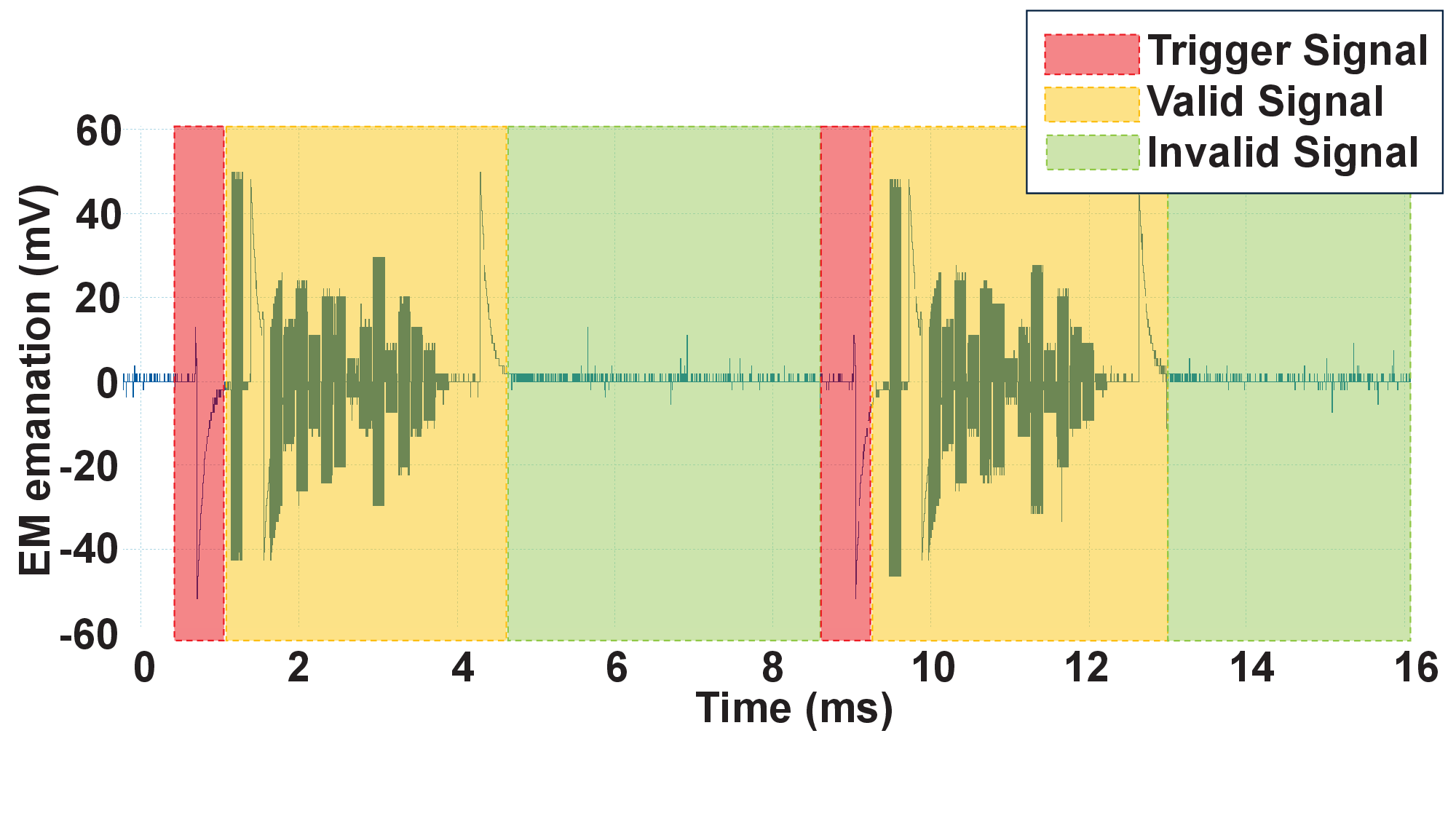}
  \vspace{-20pt}
  \caption{EM emanation measurements of two touch sampling cycles on iPhone X. The EM signal within each touch sampling cycle can be divided into three parts: trigger signal, valid signal, and invalid signal. }\label{2sample}
\end{figure}

\noindent\textbf{Attack scenario.} As illustrated in Fig. \ref{environment}, we focus on scenarios occurring in both public environments, like a library or café, and private environments, like a meeting room, where an attacker seeks to extract sensitive information from a victim using a capacitive touchscreen smartphone. These scenarios are prevalent in daily life and have been similarly considered in prior studies \cite{Li2016When, Maruyama2019Tap, jin2021periscope, Zhuoran2021Screen}. The victim user places the smartphone on a table and performs routine actions, such as entering PIN codes to unlock the device, interacting with personal applications like PayPal or WhatsApp or writing a private note. The attacker, by positioning a concealed EM probe nearby, e.g., under the table or inside a bag, as shown in Fig.~\ref{environment}, can exploit the EM signals emitted by the phone during its usage. By capturing these signals, the attacker can deduce the victim’s sensitive information without physically interacting with the phone or the victim. For example, the attacker can infer the PIN code when the victim unlocks the phone or infer the type of application the victim is using and obtain sensitive inputs, such as passwords or secret information.

\noindent\textbf{What the attacker can do.} The attacker can set up a hidden device near the victim’s phone, such as under the table or inside a book, to capture the EM signals emitted by the victim’s phone during its usage. Additionally, we assume that the attacker can identify the victim’s phone model~\cite{Chris2020How, jin2021periscope, cominelli2020even,blueID} by passively collecting information about the device brand through Bluetooth signals or network traffic, which helps determine the layout of the touchscreen and tailor the attack accordingly. Furthermore, we assume the attacker has the financial capability to purchase another smartphone of the exact same model as the victim's device to conduct offline profiling, a standard and practical assumption in modern side-channel analysis.


\noindent\textbf{What the attacker cannot do.} The attacker does not have direct access to the victim’s phone or any physical interactions with the device. This constraint poses challenges for power consumption-based attacks \cite{Spreitzer2018Systematic, Spolaor2017No}, which typically require compromised charging facilities physically connected to the target device. The attacker also cannot observe the victim’s hand or eye movements, which might otherwise provide useful clues about the input process. In addition, the attacker is unable to remotely control the victim's device via network-based attacks, spyware, or other forms of digital intrusion methods. This restriction is particularly difficult to satisfy in many sensor-based attacks \cite{oberhuber2025power, hsu2018indoor}. Lastly, the attacker does not need to know the victim's behavior during the input process, such as whether they are using a wireless charger \cite{NiZZLYWXLZ23} or performing other specific actions or physical movements.

\section{Design of \Name}
In this section, we elaborate on the design of \Name. 
Fig. \ref{Overview} provides an overview of \Name, which comprises four core phases. To address the practical challenge of acquiring ground-truth labels for neural network training, this entire pipeline operates in a two-phase workflow in real-world attacks. First, in the offline profiling phase, the attacker fully utilizes the pipeline in Fig. \ref{Overview} on an identical profiling device they control to collect labeled EM traces and train the models. Subsequently, in the online inference phase, the attacker uses the pre-trained framework to directly process unseen EM traces captured from the victim, relying on cross-device transferability without needing the victim's ground truth.




\subsection{User Interaction Classification}
The first step of \Name involves classifying user interactions based on the collected EM signals. This classification directly determines the target of the attack and influences the method for extracting sensitive information. For user touch interactions with the screen, different behaviors exhibit quite distinct characteristics in terms of single-touch duration and touch frequency. Specifically, during keyboard input, users typically demonstrate higher touch frequencies, e.g., rapid typing of longer content, and longer touch durations, e.g., quick case switching or special character input, compared to screen unlocking. While interacting with apps, users tend to touch the screen less frequently but for longer durations. Consequently, a signal acquisition period of 30 $\sim$ 60 seconds is sufficient for \Name to classify user interactions using a straightforward random forest classifier with 100 estimators and a random state of 42.

\subsection{Signal Interception}
As illustrated in Fig. \ref{2sample}, the EM signal within each touch sampling cycle exhibits distinct segmentation characteristics in the time domain. Initially, the signal shows a sharp jump, which is determined solely by the smartphone model and is unaffected by the user's touch location. Subsequently, the signal varies depending on the specific location where the user's finger touches the screen. Finally, the signal gradually diminishes to its minimum amplitude and stabilizes. Based on their impacts on \Name, we divide the signal within the touch sampling cycle into three components: the sharp jump is defined as the trigger signal; the segment containing touch location information is identified as the valid signal; and the tail signal, characterized by smooth fluctuations and small amplitude, is classified as the invalid signal.

\begin{algorithm}[t]
\caption{Signal Interception Algorithm.}
\label{alg: inter}
\begin{algorithmic}[1]
\REQUIRE Original signal $S = [s_1, s_2, \dots, s_m]$, detection window \( w \), trigger derivative threshold \( \mathcal{D}_{tri} \), trigger variance threshold $\mathcal{V}_{tri}$, termination derivative threshold \( \mathcal{D}_{ter} \), termination variance threshold $\mathcal{V}_{ter}$.
\ENSURE Valid signal $S_v$.
\STATE $F_{tri}$ $\leftarrow$ \textbf{False}, $F_{ter}$ $\leftarrow$ \textbf{False};
\FOR{$t = 1$ to $m-w$} 
    \STATE $\sigma_t^2$ $\leftarrow$ \textbf{VarianceCalculation}($S,w,t$);
    \STATE $D(t)$ $\leftarrow$ \textbf{DerivativeCalculation}($S,w,t$);
    \IF{$F_{tri} ==$ \textbf{False} \textbf{and} $D(t) > \mathcal{D}_{tri}$ \textbf{and} $\sigma_t^2 < \mathcal{V}_{tri}$}
        \STATE $t_{tri}$ $\leftarrow$ $t$, $F_{tri}$ $\leftarrow$ \textbf{True};  
    \ENDIF
    \IF{$F_{ter} ==$ \textbf{False} \textbf{and} $D(t) < \mathcal{D}_{ter}$ \textbf{and} $\sigma_t^2 < \mathcal{V}_{ter}$}
        \STATE $t_{ter}$ $\leftarrow$ $t$, $F_{ter}$ $\leftarrow$ \textbf{True};  
    \ENDIF
\ENDFOR
\STATE $S_v$ $\leftarrow$ \textbf{SignalInterception}($S,t_{tri},t_{ter}$); 
\RETURN $S_v$;
\end{algorithmic}
\end{algorithm}

Consequently, we develop a signal interception algorithm to enhance the signal-to-noise ratio. As shown in Algorithm \ref{alg: inter}, the signal interception algorithm establishes a detection window of length $w$ and calculates the first derivative $D(t)$ and variance $\sigma^2_t$ point by point from the starting position (Lines 3-4). Subsequently, these calculations are compared with ($\mathcal{D}_{tri}$, $\mathcal{V}_{tri}$) for trigger detection and ($\mathcal{D}_{ter}$, $\mathcal{V}_{ter}$) for termination detection (Lines 5 and 8). When the derivative $D(t)$ exceeds the threshold $\mathcal{D}_{tri}$ and the variance $\sigma^2_t$ is below the threshold $\mathcal{V}_{tri}$, i.e., when the signal exhibits a sudden sharp jump from a gentle state, the signal within the window is identified as the trigger signal and serves as the starting point $t_{tri}$ for signal interception (Line 6). When the derivative $D(t)$ falls below the threshold $\mathcal{D}_{ter}$ and the variance $\sigma^2_t$ is below the threshold $\mathcal{V}_{ter}$, i.e., when the signal stabilizes, the signal within the window is identified as the termination of the valid signal and serves as the endpoint $t_{ter}$ for signal interception (Line 9). Finally, the original signal $S$ is intercepted from the start point $t_{tri}$ to the endpoint $t_{ter}$ to extract the valid signal $S_v$ (Line 12).

\subsection{Signal Normalization}
After valid signals are intercepted, given that the lengths of these signals may vary and amplitude fluctuations can be caused by ambient noise in real scenarios as well as phone movement during touch interactions, \Name subsequently performs signal normalization to ensure optimal inputs for sensitive information extraction.

As shown in Algorithm 2, the intercepted signal $S_v$ is first reshaped to the target length $L$. For signals shorter than $L$, we increase the number of data points through up-sampling (Lines 1-2). Conversely, for signals longer than $L$, we reduce the number of data points through down-sampling (Lines 3-4). After unifying the signal length, we normalize the signal amplitude using $Z$-score normalization (Line 6), ensuring that all signals have a mean of 0 and a standard deviation of 1.

\begin{algorithm}[!tbp]
\caption{Signal Normalization Algorithm.}
\label{alg: normal}
\begin{algorithmic}[1]
\REQUIRE The intercepted signal \( S_v\), target length \( L \);
\ENSURE The normalized signals \( {S}_{n}\).
\IF{ \( \textbf{Len}(S_{v}) < L \)} 
    \STATE $S_{n}$ $\leftarrow$ \textbf{SignalUpSampling}($S_{v},\textbf{Len}(S_{v}),L$);
\ELSIF{ \( \textbf{Len}(S_{v}) > L \)} 
    \STATE $S_{n}$ $\leftarrow$ \textbf{SignalDownSampling}($S_{v},\textbf{Len}(S_{v}),L$);
\ENDIF
\STATE $S_{n}$ $\leftarrow$ \textbf{ZscoreNormalization}($S_n,L$) ;
\RETURN ${S}_{n}$;
\end{algorithmic}
\end{algorithm}

\subsection{Sensitive Information Extraction}
Following the signal normalization phase, \Name employs deep learning classifiers to extract sensitive information from the processed signal. For each distinct attack target, \Name develops a specialized deep neural network architecture aimed at extracting sensitive information with high precision and reliability. 

In screen-unlock PIN code recovery and keyboard input recovery scenarios, the core of sensitive information extraction lies in recognizing touch interactions across different regions of the touchscreen. Consequently, for PIN code recovery, the touchscreen is segmented into 10 distinct regions corresponding to digits 0-9. For keyboard input recovery, the touchscreen is divided into finer regions to align with the letters on a virtual keyboard and a MLP classifier is employed for achieving accurate recovery. The selection of a lightweight network architecture (e.g., MLP or CNN), as opposed to complex sequence models (e.g., LSTMs), is a deliberate design choice driven by the preceding module. Specifically, our signal interception and normalization phases strictly synchronize the temporal features of isolated taps into fixed-length, position-aligned vectors. This robust alignment essentially eliminates the need to track long-range temporal dependencies within a single discrete touch trace. Consequently, a lightweight model is highly sufficient to extract local distinguishing features, significantly reducing the computational overhead for the attacker without sacrificing accuracy.

\begin{figure*}[!htbp]
  \centering
  \subfloat[]{\includegraphics[width=.23\linewidth]{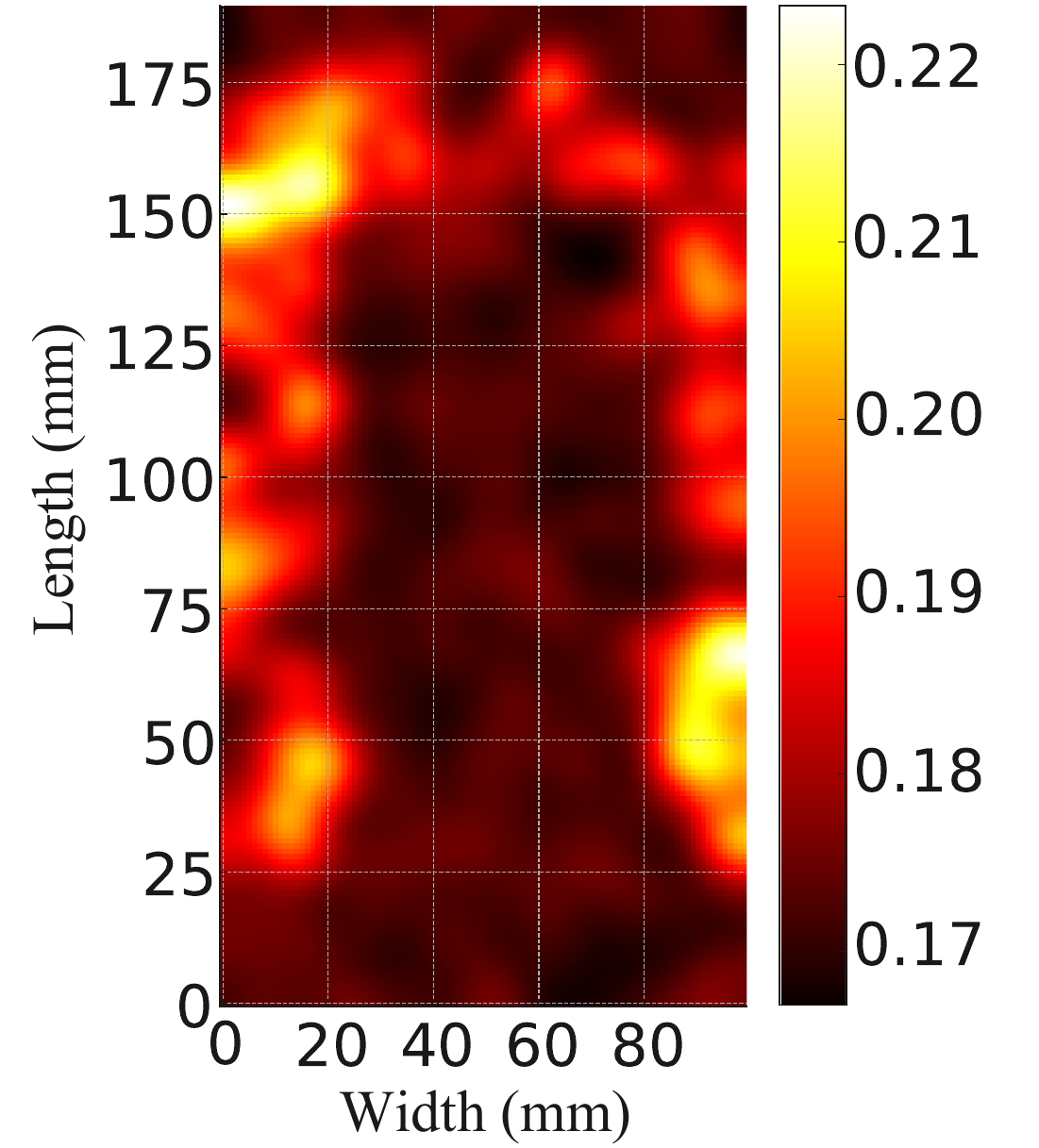}}%
  \subfloat[]{\includegraphics[width=.23\linewidth]{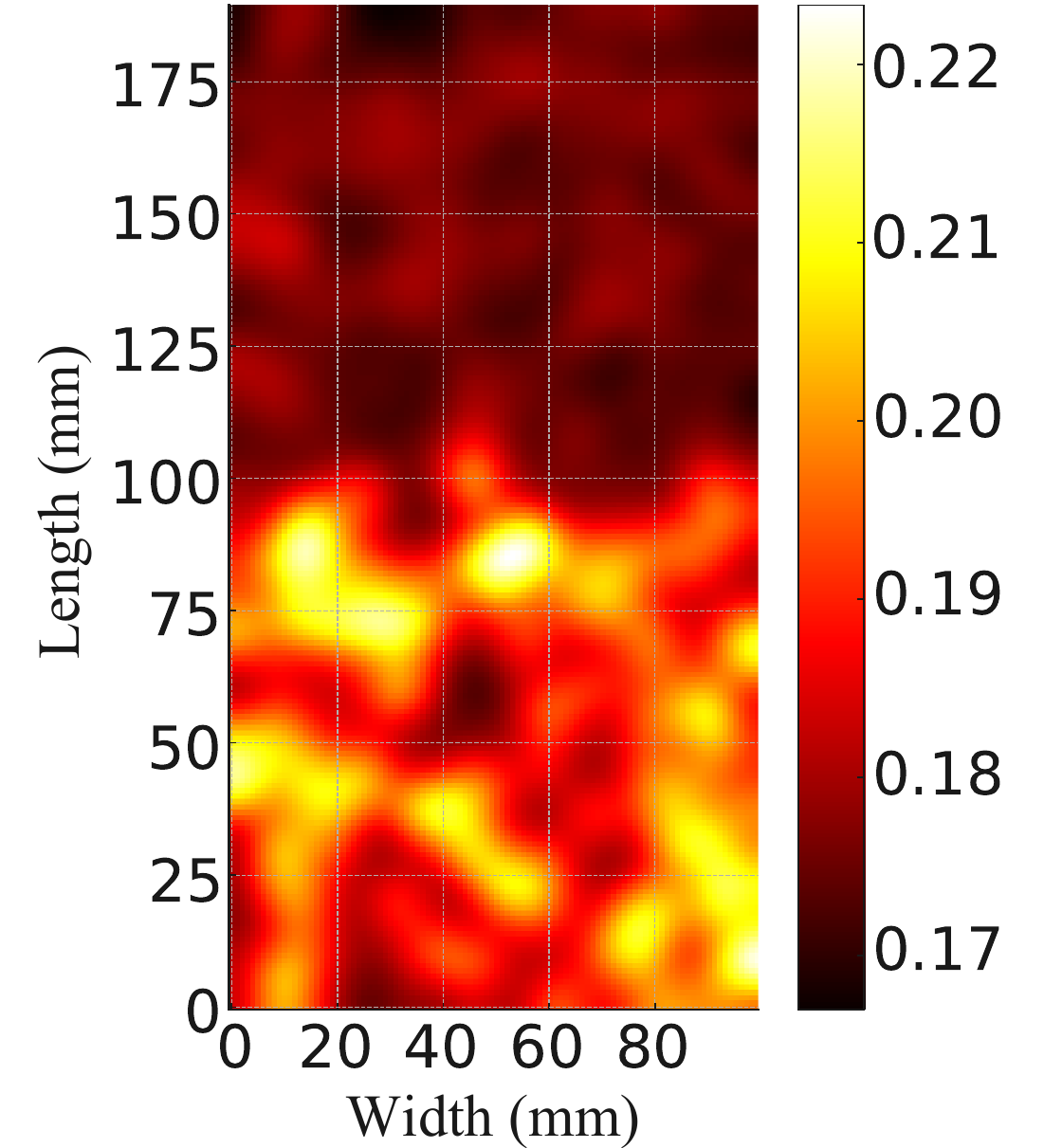}}
  \subfloat[]{\includegraphics[width=.23\linewidth]{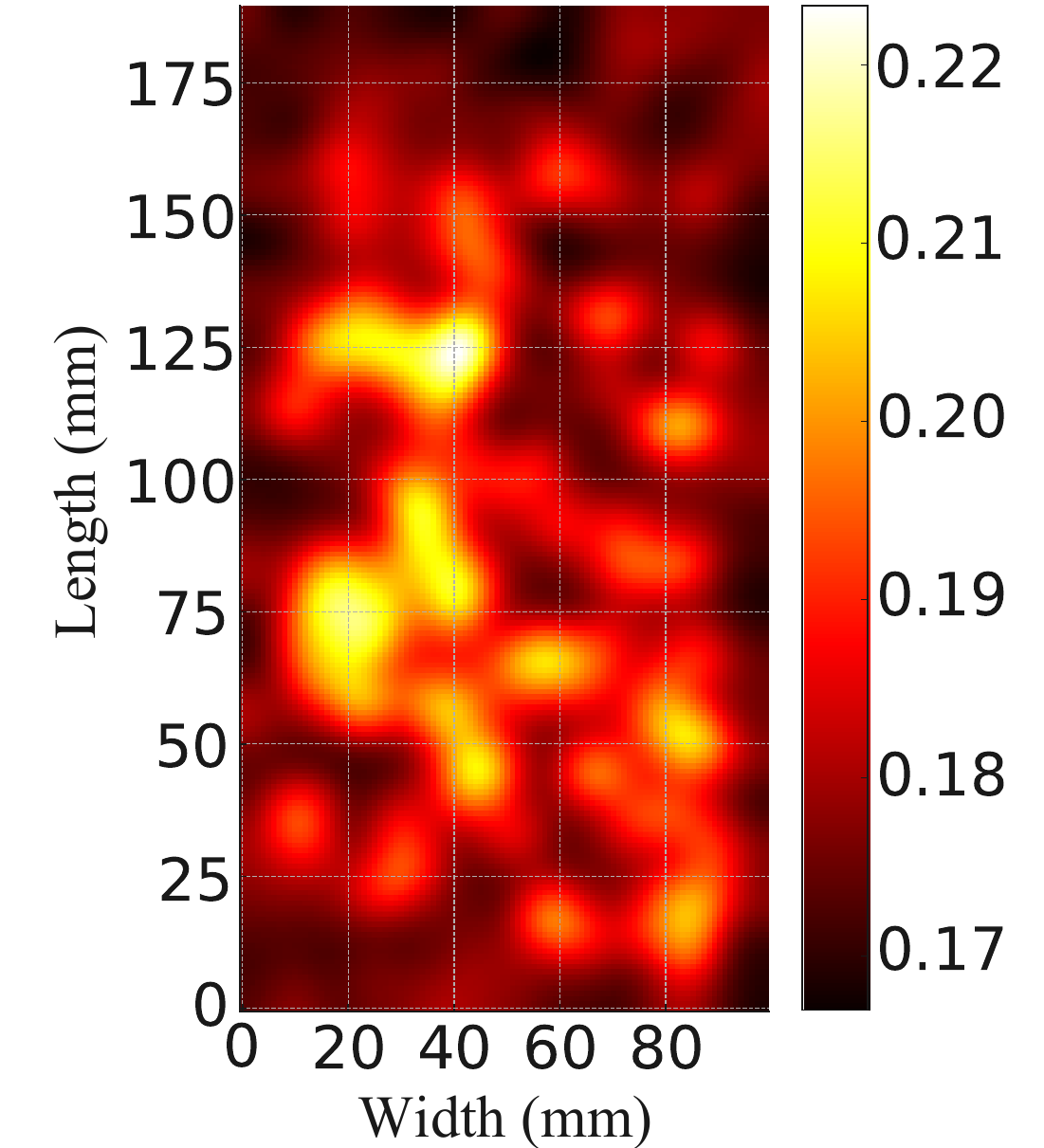}}
  \subfloat[]{\includegraphics[width=.23\linewidth]{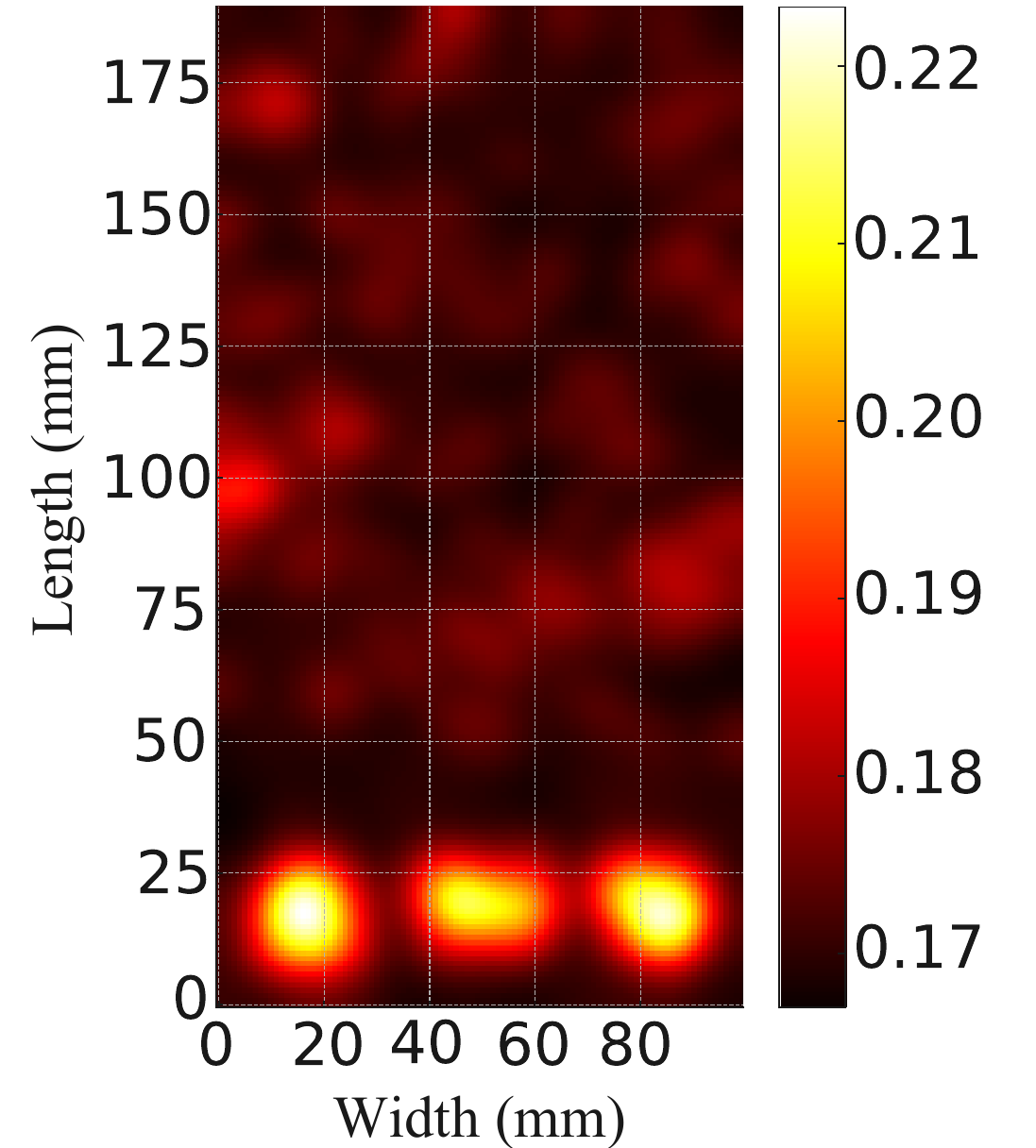}}
  \caption{Touch heatmap for distinct application categories. (a) App Store application, (b) Mobile payment application, (c) Shopping application, (d) Music application.}\label{heatmap}
\end{figure*}

Moreover, \Name also aims to recover the interacting application categories. This recovery enables attackers to identify the specific application categories being used by the victim, thereby determining the significance of the recovered sensitive information.  For this more complex recovery target, \Name extracts sensitive information via touch hotspots. Specifically, as illustrated in Fig. \ref{heatmap}, different application categories demonstrate distinct tendencies in touch zone distribution: (1) app store applications concentrate their touch hotspots primarily at the top and both sides of the screen for operations such as application search, icon selection, and installation; (2) mobile payment applications focus their touch hotspots mainly in the lower half of the screen for tasks like entering amounts and passwords; (3) shopping applications cluster their touch hotspots predominantly in the central area of the screen for browsing product images; (4) music applications localize their touch hotspots largely at the bottom of the screen for actions including playing, pausing, and switching songs. Consequently, a CNN classifier is employed after the touch location recovery to address this more intricate scenario.

Beyond discrete touch inputs and macroscopic application interactions, \Name further extends its capability to reconstruct continuous, fine-grained handwriting trajectories. This represents a highly challenging scenario as it requires mapping a continuous signal stream directly to dynamic two-dimensional (2D) spatial coordinates. To achieve this, the trajectory reconstruction process is structured into a three-step pipeline: position recovery, post-processing, and splicing. First, in the position recovery step, the processed and normalized EM signals are sequentially fed into a transformer-based deep neural network. This network is specifically trained to capture the temporal dependencies in the continuous signal, estimating a sequence of touch positions and producing 2D coordinates for each time step. Next, the raw coordinate sequence undergoes post-processing, during which smoothing techniques, such as filtering and interpolation, are applied to reduce noise and mitigate jitter introduced during the initial recovery phase. Finally, in the splicing step, the refined positional estimates are connected in their correct temporal order to generate the final, clean reconstructed handwritten trajectory.

\begin{figure}[!tbp]
  \centering
  \includegraphics[width=0.65\linewidth]{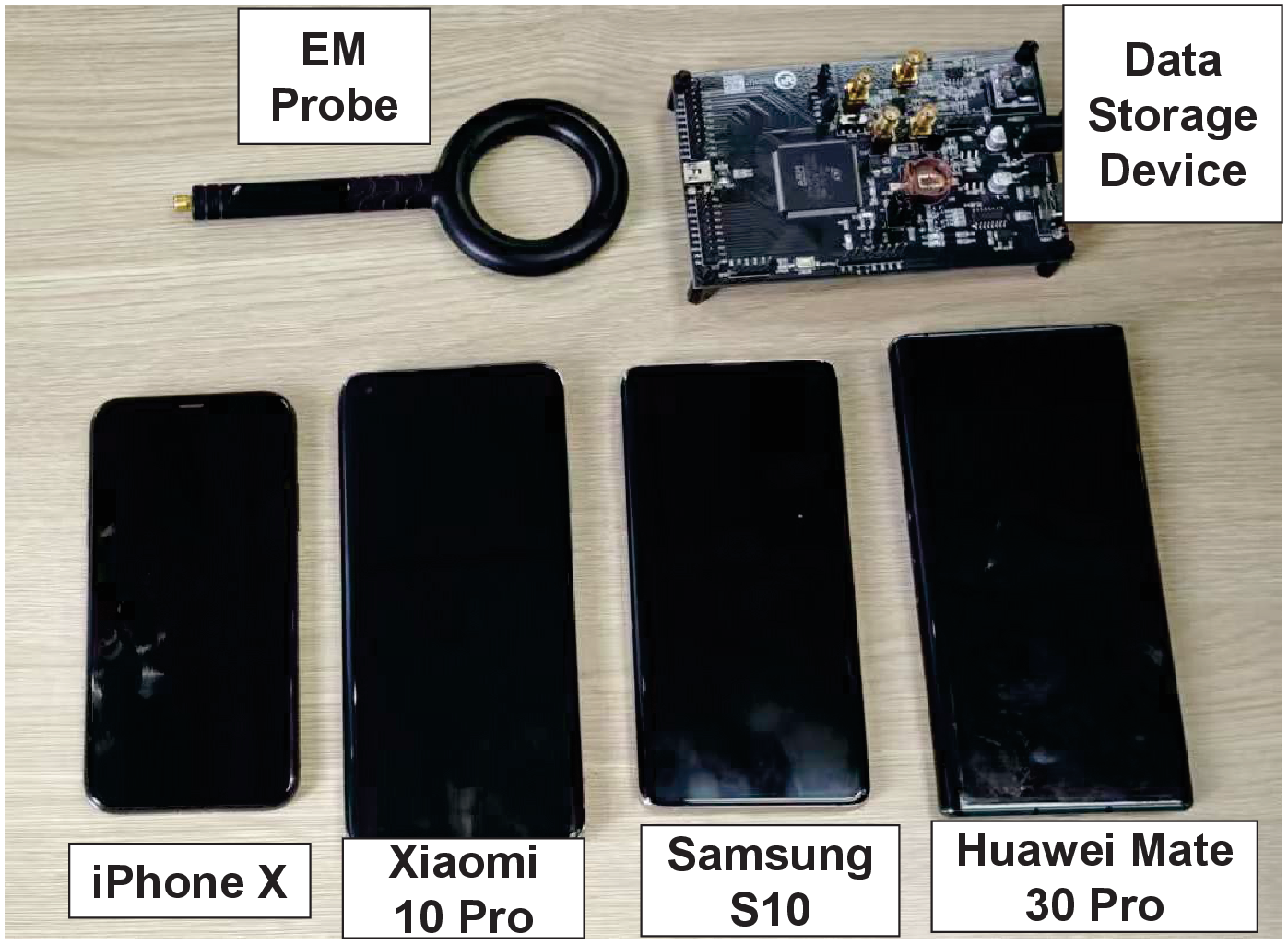}
  \caption{The target smartphones and data collection devices utilized in the evaluation.}\label{device}
\end{figure}

\section{Evaluations}
\label{sec: evaluation}

\subsection{Experimental Setups}
To account for individual variability in smartphone interaction patterns, e.g., preferred touch areas and durations, we recruited a diverse group of 20 participants (16 males and 4 females, aged 22–27 years) for data collection. Each participant was placed in similar environments (e.g., different meeting rooms with analogous layouts or different seats on the same floor of the library) and asked to perform three representative interaction tasks: (i) screen unlocking using randomly generated PIN codes, (ii) text input via the virtual keyboard, (iii) interacting with various applications, and (iv) handwritting random characters and words. The smartphone was naturally placed on the table, allowing participants to interact with it according to their habitual behavioral patterns while permitting slight movements of the device. To rigorously evaluate \Name's performance and prevent overfitting, we adopt an 80:20 randomized train-test split across the entire dataset.

For data collection, four commercial smartphones, i.e.,  iPhone X, Xiaomi 10 Pro, Samsung S10, and Huawei Mate 30 Pro, were used as the victim devices. We first analyzed the attack performance of \Name on iPhone X and subsequently presented the results for the other phones. An EM measurement probe was placed at a fixed distance of 5 cm directly facing the device surface. To ensure controlled experimental conditions, all background third-party applications were terminated while essential system services that maintain core device functionality were preserved. Our data collection system consisted of several components: a Fosttek NFP-One P1 EM measurement probe, 
which captures EM signals at a sampling frequency of 500 kHz, and an STM32F407 microcontroller unit (MCU) for data sampling and storage. 
The target smartphones and data collection devices utilized in the evaluation are presented in Fig. \ref{device}.  Additionally, a Picoscope 5444D oscilloscope was used as an optional tool for signal observation and validation.

The data processing and analysis were performed on a high-performance Dell Precision server equipped with an Intel Xeon Silver 4210R CPU running at 2.40 GHz, 32 GB of DDR memory, a 1 TB Samsung SSD-980-PRO storage drive, and an NVIDIA GeForce RTX 3090 GPU. Our software environment utilized Python 3.9.7 as the primary development language within Jupyter Notebook 6.4.5. The deep learning implementation was built using Keras 2.9.0 as the frontend with TensorFlow 2.9.1 as the backend, leveraging CUDA 11.7 for GPU acceleration during model training. The training duration for each model ranges from approximately 2 to 5 minutes, while the evaluation time spans from several seconds to tens of seconds.

\subsection{Deep Learning Model Configurations}
As presented in Table \ref{MLP}, the Multi-Layer Perceptron (MLP) utilized in our evaluations is a fully connected neural network consisting of four fully connected layers (FC). Each FC layer is followed by batch normalization and dropout layers for regularization. The input is a vector with $n_{input}$ features, where the number of features is determined by the model of the target smartphone and processed through three hidden layers. The first fully connected layer (FC 1) contains 512 neurons and employs the ReLU activation function, followed by batch normalization (BatchNorm 1) to normalize the output and mitigate internal covariate shift. A dropout rate of 30\% is applied to prevent overfitting. The second fully connected layer (FC 2) reduces the number of neurons to 256, followed by another batch normalization (BatchNorm 2) and a dropout layer. The third fully connected layer (FC 3) further reduces the number of neurons to 128, with batch normalization (BatchNorm 3) and a dropout rate of 20\%. The final fully connected layer (FC 4) outputs a vector of size $n_{class}$, corresponding to the number of classes determined by the type of sensitive information, e.g., 10 for PIN codes or 26 for characters. Additionally, the MLP utilizes the Adam optimizer with an initial learning rate of 0.001.

\begin{table}[!tbp]
\centering
\caption{MLP model configurations.}
\label{MLP}
\setlength{\tabcolsep}{4.5mm}{\resizebox{\linewidth}{!}{
\begin{tabular}{c|c|c|c}
\Xhline{1.5pt}
\textbf{Layer}                     & \textbf{Output Shape}   & \textbf{Activation} & \textbf{Parameters} \\ \Xhline{1.5pt}
\textbf{Input}                      & (1, $n_{input}$)               & -                    & 0                    \\ \hline
\textbf{Flatten}                    & (1, 2150)               & -                    & 0                    \\ \hline
\textbf{FC 1} & (1, 512)                & ReLU                 & 1,102,080            \\ \hline
\textbf{BatchNorm 1}               & (1, 512)                & -                    & 1,024                \\ \hline
\textbf{Dropout(0.3)}                  & (1, 512)                & -                    & 0                    \\ \hline
\textbf{FC 2} & (1, 256)                & ReLU                 & 131,328              \\ \hline
\textbf{BatchNorm 2}               & (1, 256)                & -                    & 512                  \\ \hline
\textbf{Dropout(0.3)}                  & (1, 256)                & -                    & 0                    \\ \hline
\textbf{FC 3} & (1, 128)                & ReLU                 & 32,896               \\ \hline
\textbf{BatchNorm 3}               & (1, 128)                & -                    & 256                  \\ \hline
\textbf{Dropout(0.2)}                  & (1, 128)                & -                    & 0                    \\ \hline
\textbf{FC 4} & (1, $n_{class}$)                 & -                    & 1,290                \\ \Xhline{1.5pt}
\end{tabular}}}
\end{table}

As shown in Table \ref{CNN}, the Convolutional Neural Network (CNN) consists of two convolutional layers, each followed by max-pooling layers, a flattening layer, and two fully connected layers, making it suitable for extracting hierarchical features from sequential data. The length of the input depends on the model of the target smartphone. The first convolutional layer (Conv 1) applies 16 filters, each of size 8 with a stride of 4, producing 16 feature maps of reduced length. A ReLU activation function is then applied to introduce non-linearity. The first max-pooling (MaxPool 1) layer, with a kernel size of 4, downsamples the feature maps, further reducing their length. The second convolutional layer (Conv 2) applies 32 filters, also of size 8 with a stride of 4, followed by another max-pooling layer (MaxPool 2) that further reduces the feature map length. The resulting feature maps are flattened into a 1D vector of size 320 and passed through two fully connected layers. The first fully connected layer  (FC 1) has 128 neurons, followed by a ReLU activation function and a 50\% dropout layer to prevent overfitting. The final fully connected layer  (FC 2) outputs a vector of size 4, representing the number of application categories evaluated, i.e., app store, mobile payment, shopping, and music. The CNN also utilizes the Adam optimizer with a larger initial learning rate of 0.01.

\begin{table}[!tbp]
\centering
\caption{CNN model configurations.}
\label{CNN}
\resizebox{\linewidth}{!}{
\begin{tabular}{c|c|c|c|c|c|c}
\Xhline{1.5pt}
\textbf{Layer}                     & \textbf{Output Shape}   & \textbf{Kernel Size} & \textbf{Stride} & \textbf{Padding} & \textbf{Activation} & \textbf{Parameters} \\ \Xhline{1.5pt}
\textbf{Input}                      & (1, 1, $n_{input}$)             & -                     & -                & -                 & -                    & 0                    \\ \hline
\textbf{Conv 1}                      & (1, 16, 198)            & 8                     & 4                & 0                 & ReLU                 & 144                  \\ \hline
\textbf{MaxPool 1}                   & (1, 16, 49)             & 4                     & 4                & 0                 & -                    & 0                    \\ \hline
\textbf{Conv 2}                      & (1, 32, 42)             & 8                     & 4                & 0                 & ReLU                 & 4,160                \\ \hline
\textbf{MaxPool 2}                   & (1, 32, 10)             & 4                     & 4                & 0                 & -                    & 0                    \\ \hline
\textbf{Flatten}                    & (1, 320)                & -                     & -                & -                 & -                    & 0                    \\ \hline
\textbf{FC 1}    & (1, 128)                & -                     & -                & -                 & ReLU                 & 41,088               \\ \hline
\textbf{Dropout}                    & (1, 128)                & -                     & -                & -                 & -                    & 0                    \\ \hline
\textbf{FC 2}    & (1, 4)                  & -                     & -                & -                 & -                    & 516                  \\ \Xhline{1.5pt}
\end{tabular}}
\end{table}

The architecture of our transformer model is detailed in Table \ref{table: trans}. It begins with a 1D convolutional (Conv1D) front-end for feature extraction, followed by a Transformer Encoder that processes the resulting sequence with a model dimension ($d_{model}$) of 256, a feed-forward dimension ($d_{ff}$) of 1024, and a maximum sequence length ($max\_len$) of 5000. An attention pooling layer aggregates the encoder outputs, which are then passed through a multi-layer perceptron (MLP) composed of three fully connected layers (FC) for signal classification.

\begin{table}[t]
\centering
\caption{Transformer model architecture.}
\renewcommand{\arraystretch}{1}
\resizebox{\linewidth}{!}{
\begin{tabular}{c|c|c}
\Xhline{1.5pt}
\textbf{Layer Group}   & \textbf{Output Shape} & \textbf{Key Parameters / Structure}   \\ \Xhline{1.5pt}
Input                 &  (1, $N_{input}$)            &                     -                                                                                            \\ \hline
Convolutional Front-End & (256, 445)            & \begin{tabular}[c]{@{}c@{}}Conv1d(1, 64, kernel=7, stride=2) -\textgreater \\ Conv1d(64, 256, kernel=5, stride=2)\end{tabular} \\ \hline
Positional Encoding    & (445, 256)            & $d_{model}$=256, $max\_len$=5000                                                                             \\ \hline
Transformer Encoder    & (445, 256)            &  $d_{ff}$=1024                                                                                                        \\ \hline
Attention Pool         & (256)                 & FC(256-\textgreater{}128) -\textgreater FC(128-\textgreater{}1)\\ \hline
Classifier (MLP)       & ($N_{class}$)            & FC(256-\textgreater{}512) -\textgreater FC(512-\textgreater{}256) -\textgreater FC(256-\textgreater{}$N_{class}$)                                                                                      \\ \Xhline{1.5pt}
\end{tabular}}
\label{table: trans}
\end{table}

\subsection{User Interaction Classification}

Our evaluation starts with user interaction classification. For each interaction modality (screen unlocking, keyboard input, and application operation), we collected continuous 30-second EM trace segments that represent complete instances of user interactions. This data acquisition process was repeated 500 times per behavior type, resulting in a comprehensive dataset comprising 1,500 samples. We partitioned the dataset into training and testing subsets at an randomized 80:20 ratio (1,200 training samples and 300 testing samples) to ensure a representative distribution of interaction patterns while avoiding data leakage between the training and evaluation sets.

\begin{wrapfigure}{r}{0.5\linewidth}
  \vspace{-15pt}
  \begin{center}
    \includegraphics[width=0.88\linewidth]{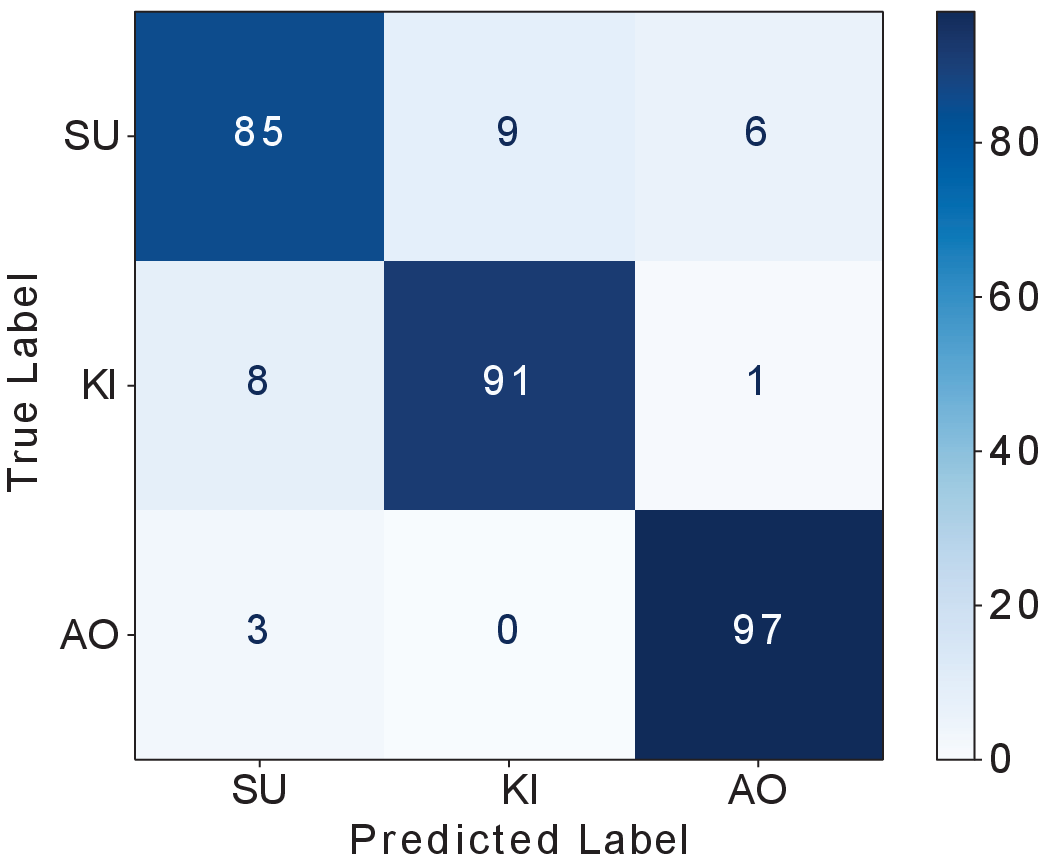}
  \end{center}
  \vspace{-5pt}
  \caption{Effectiveness on interaction classification. SU: screen unlocking; KI: keyboard input; AO: application operation.}
  \vspace{-5pt}
  \label{3State}
\end{wrapfigure} 

Fig. \ref{3State} demonstrates that our classifier attains an overall accuracy of 91.0\% in distinguishing among three distinct types of user interactions. Notably, higher misclassification rates are observed between screen unlocking and keyboard input compared to other class pairs. This is likely due to the inherent similarity in their interaction patterns—both modalities involve frequent touch inputs during initial device engagement, whereas application operations typically exhibit more distinctive and sustained interaction features.


\subsection{Sensitive Information Recovery}
\label{sec: recovery}
\noindent\textbf{Screen-Unlocking PIN Codes Recovery.} 
To evaluate the effectiveness of recovering screen-unlocking PIN codes, we constructed a targeted dataset by collecting EM traces for digits 0-9 PIN codes across various screen locations. We repeated the data acquisition process 1,200 times per digit, resulting in a total of 12,000 samples that comprehensively capture spatial variations during PIN entry interactions. These samples were then divided into training and testing subsets at a ratio of 80:20, yielding 9,600 training samples and 2,400 testing samples.

\begin{figure}[!tbp]
  \centering
  \includegraphics[width=0.9\linewidth]{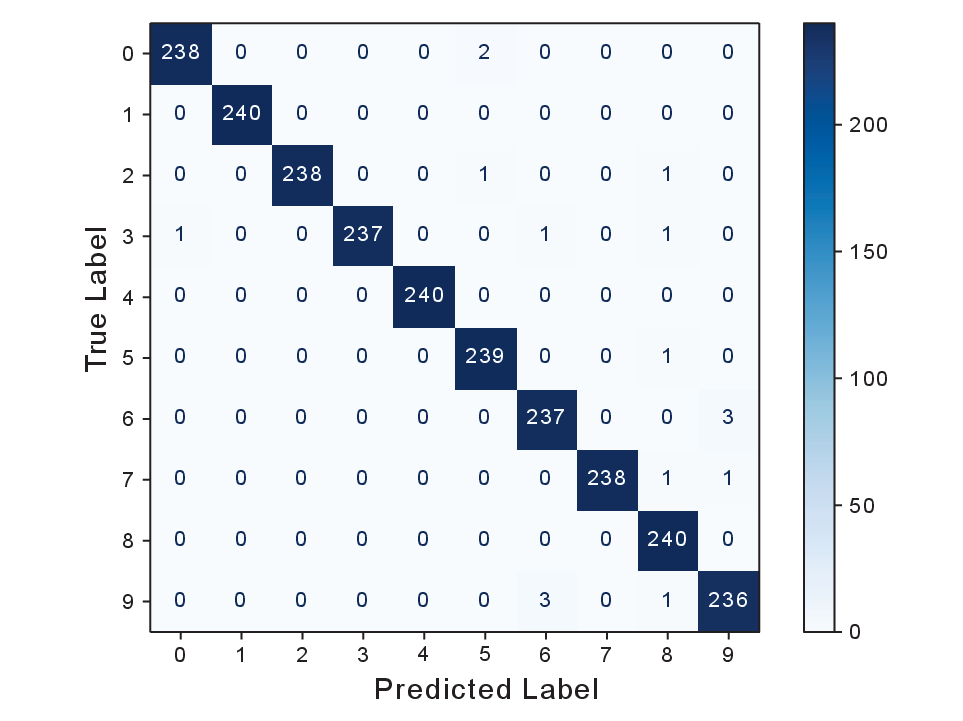}
  \caption{Effectiveness evaluation on screen-unlocking PIN codes recovery.}\label{iPhoneX_PIN}
\end{figure}

The evaluation results, as depicted in the confusion matrix of Fig. \ref{iPhoneX_PIN}, achieve an exceptional overall classification accuracy of 99.3\% across digit classes. Notably, residual misclassification primarily occurs between digits within the same column, exhibiting direct correlation with our analysis findings in Section \ref{sec: touch location}, where vertical position variations demonstrate smaller EM signature disparities compared to horizontal position differences. This error pattern stems from the touchscreen's scan-driven architecture: vertically aligned digits share common TX electrode activation patterns during sampling cycles, resulting in more similar EM emission profiles than horizontally separated digits. 

\noindent\textbf{Keyboard Inputs Recovery.} 
Building upon the successful framework established for PIN code recovery, we maintained consistent experimental conditions while extending the data collection to address the more challenging keyboard input scenario. The target dataset was constructed for all 26 alphabetic characters on the virtual keyboard, consisting of 1,200 EM trace samples per character and preserving the same 80:20 training-to-testing ratio used in the experiments.

\begin{figure}[!tbp]
  \centering
  \includegraphics[width=0.98\linewidth]{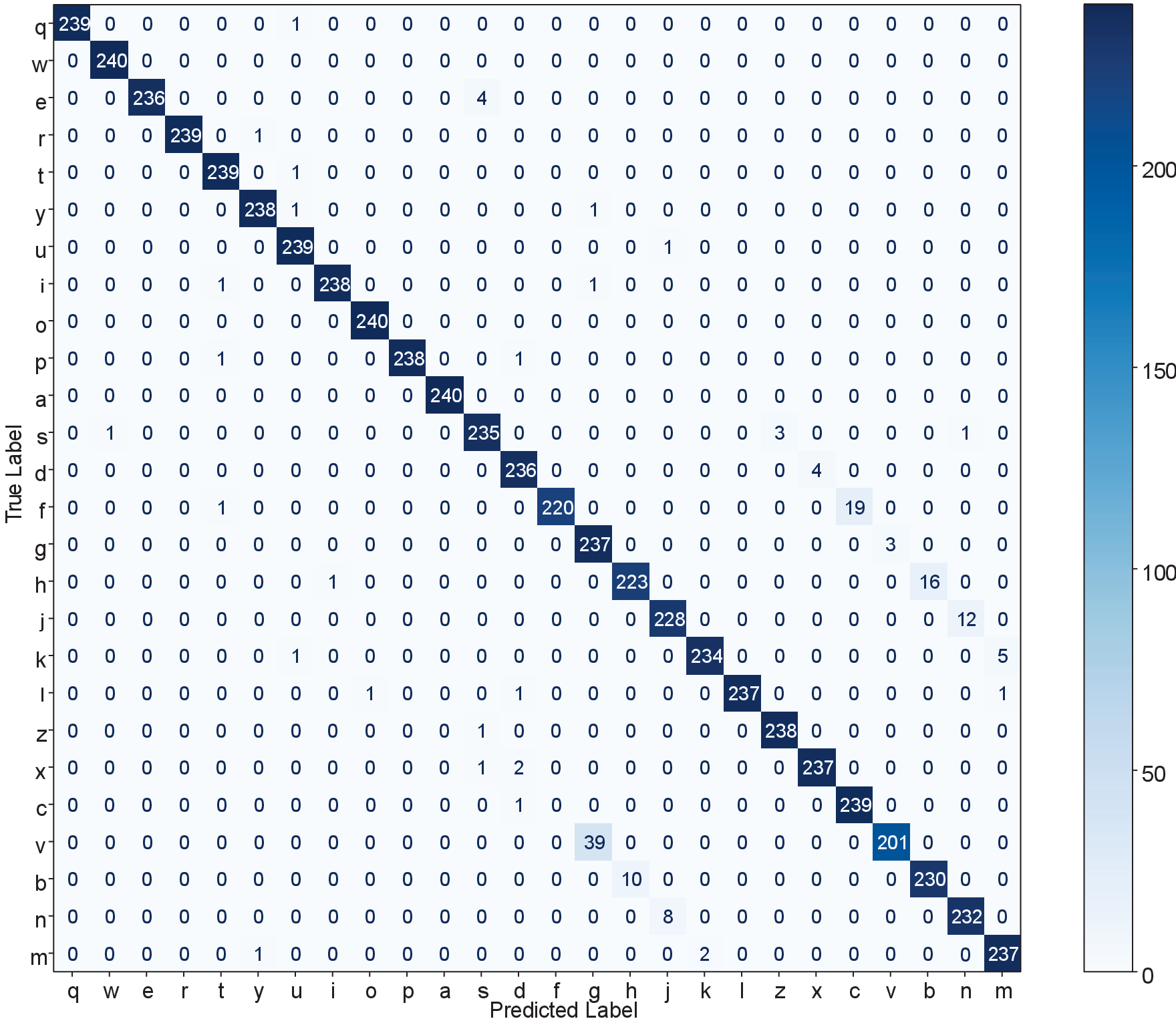}
  \caption{Effectiveness evaluation on keyboard inputs recovery.}\label{iPhoneX_keyboard}
\end{figure}

As shown in Fig. \ref{iPhoneX_keyboard}, our keyboard input recovery attack achieves an overall accuracy of 97.6\%. An analysis of the error cases reveals that the majority of misclassifications occur between vertically adjacent letters on the QWERTY keyboard layout, specifically: `f' being misidentified as `c', `h' as `n', `v' as `g', and `b' as `h'. These results further corroborate our hypothesis regarding the EM feature similarity of vertically aligned keys.

\noindent\textbf{Application Categories Recovery.} 
We evaluate the performance of \Name in recovering application categories by comparing the touch hotspots of distinct applications. Specifically, we selected representative apps for each category: Apple App Store and Google Play Store for App Store, PayPal and Alipay for Mobile Payment, Amazon and Tmall for Shopping, Spotify and QQ music for Music. To construct the dataset for application category recovery, we collected samples by identifying the distribution of users' multiple touch interactions on the touchscreen over a continuous 300-second EM trace. For each application category, the target dataset consisted of 75 samples and followed the same 80:20 training-to-testing ratio.

\begin{figure}[!tbp]
  \centering
  \includegraphics[width=0.72\linewidth]{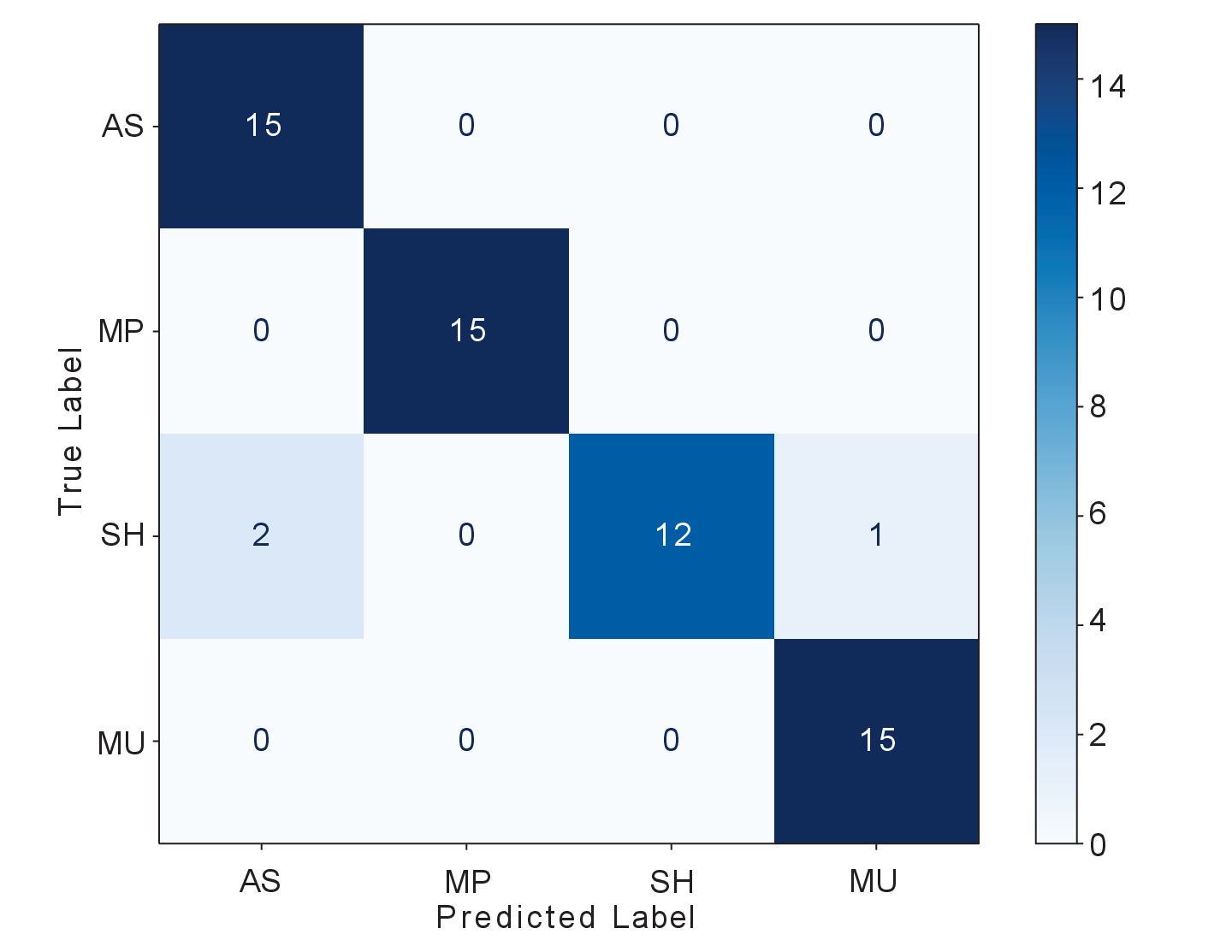}
  \caption{Effectiveness evaluation on application categories recovery. AS: app store; MP: mobile payment; SH: shopping; MU: music.}\label{APP}
  \vspace{-5pt}
\end{figure}

Fig. \ref{APP} presents the evaluation results of application category recovery, achieving an overall accuracy of 95.0\%. Misclassifications predominantly occur in the classification of shopping applications due to their more evenly distributed touch hotspots and less distinct characteristics compared to the other three categories.

\begin{figure}[!tbp]
  \centering
  \includegraphics[width=\linewidth]{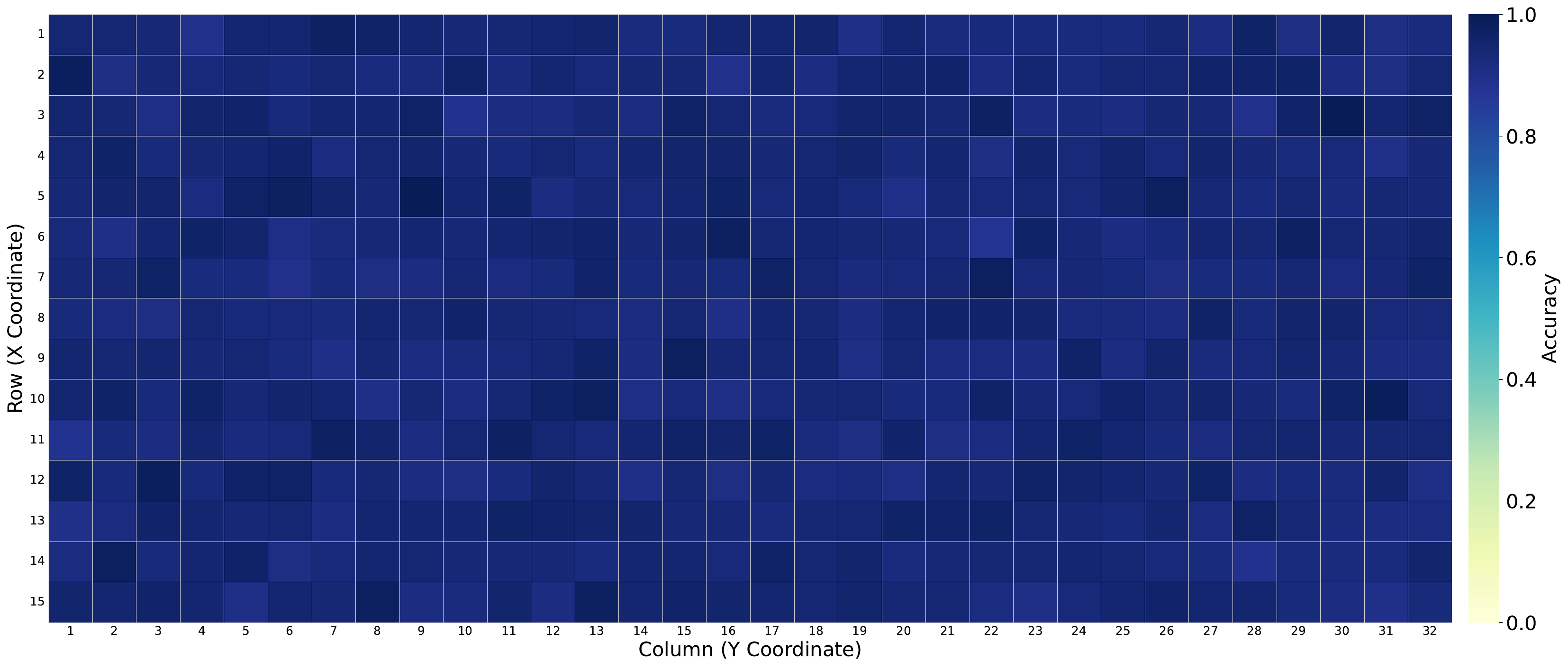}
  \caption{Evaluation results of touch position recovery.}\label{fig: accuracy}
\end{figure}

\noindent\textbf{Handwriting Trajectory Recovery.} 
Beyond discrete inputs, we evaluated the performance of \Name in recovering continuous handwriting trajectories. The evaluation was conducted progressively, starting from fundamental static touch position recovery to complex character and word-level trajectory reconstructions. First, to assess the model's baseline spatial resolution, we divided the smartphone screen into a 32$\times$15 grid, resulting in 480 distinct zones. We collected a comprehensive dataset of 480,000 samples across the screen (i.e., 1000 samples per zone). As shown in Figure \ref{fig: accuracy}, \Name achieved an exceptional overall accuracy of 94.08\% in identifying the correct grid region, demonstrating its capability for high-resolution spatial tracking.

\begin{figure}[t]
\centering
\subfloat[]{\includegraphics[width=.5\linewidth]{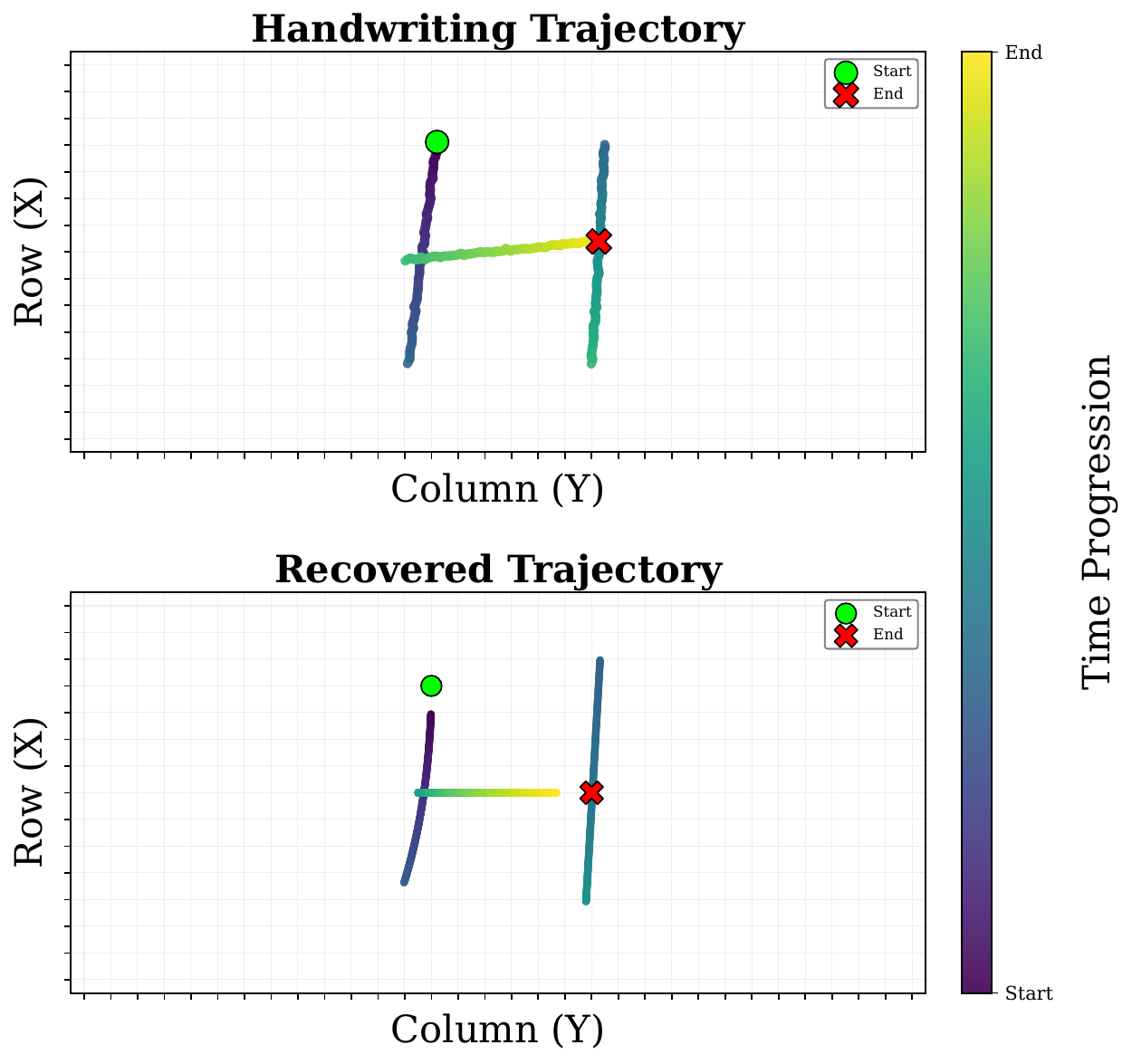}\label{fig: letter}}%
\subfloat[]{\includegraphics[width=.5\linewidth]{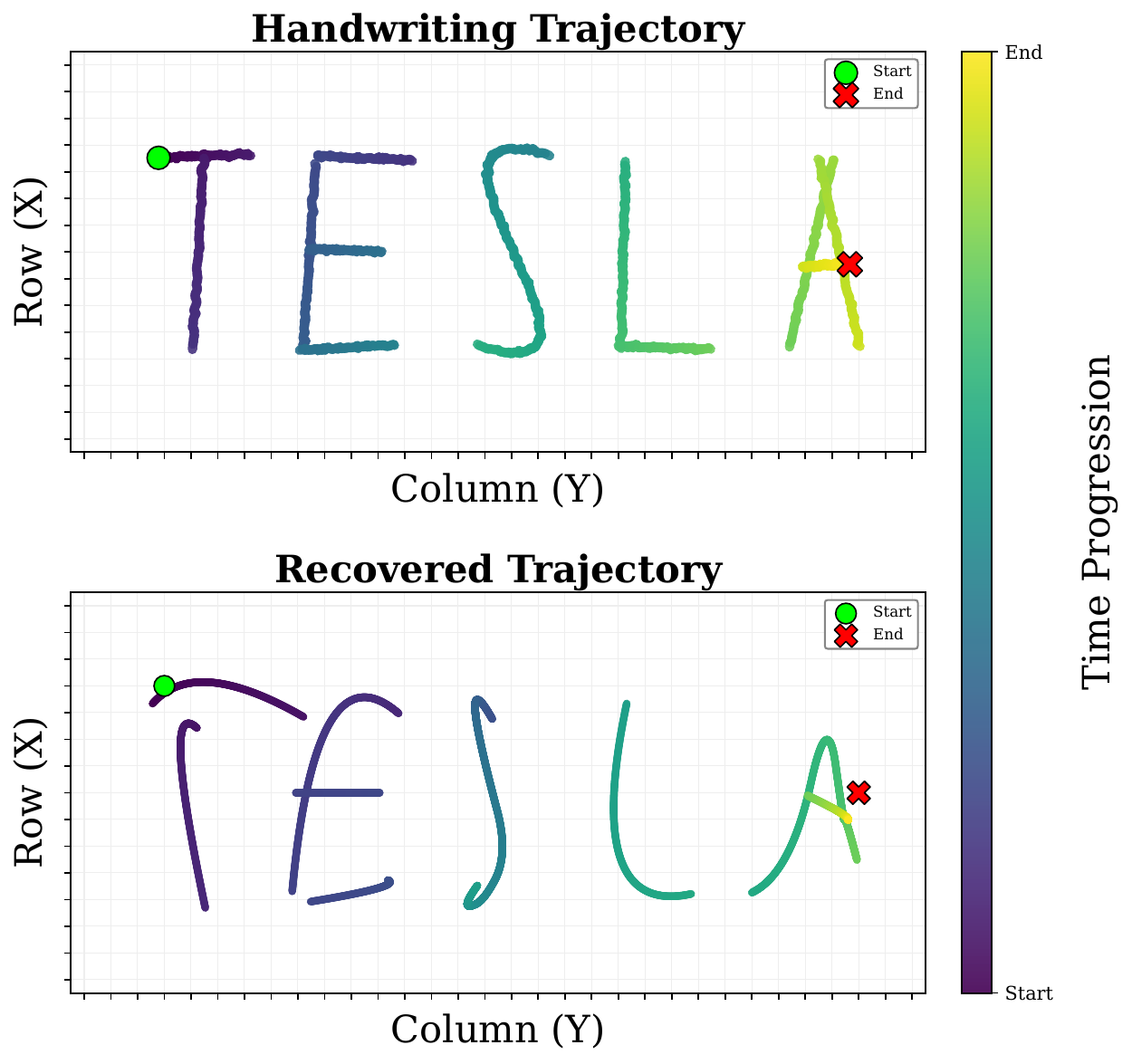}\label{fig: word}}
\vspace{-10pt}
\caption{Comparisons of handwriting trajectory and recovered trajectory. (a) Character-level. (b) Word-level.}
\label{fig: handwriting}
\end{figure}

\begin{figure}[!tbp]
  \centering
  \includegraphics[width=\linewidth]{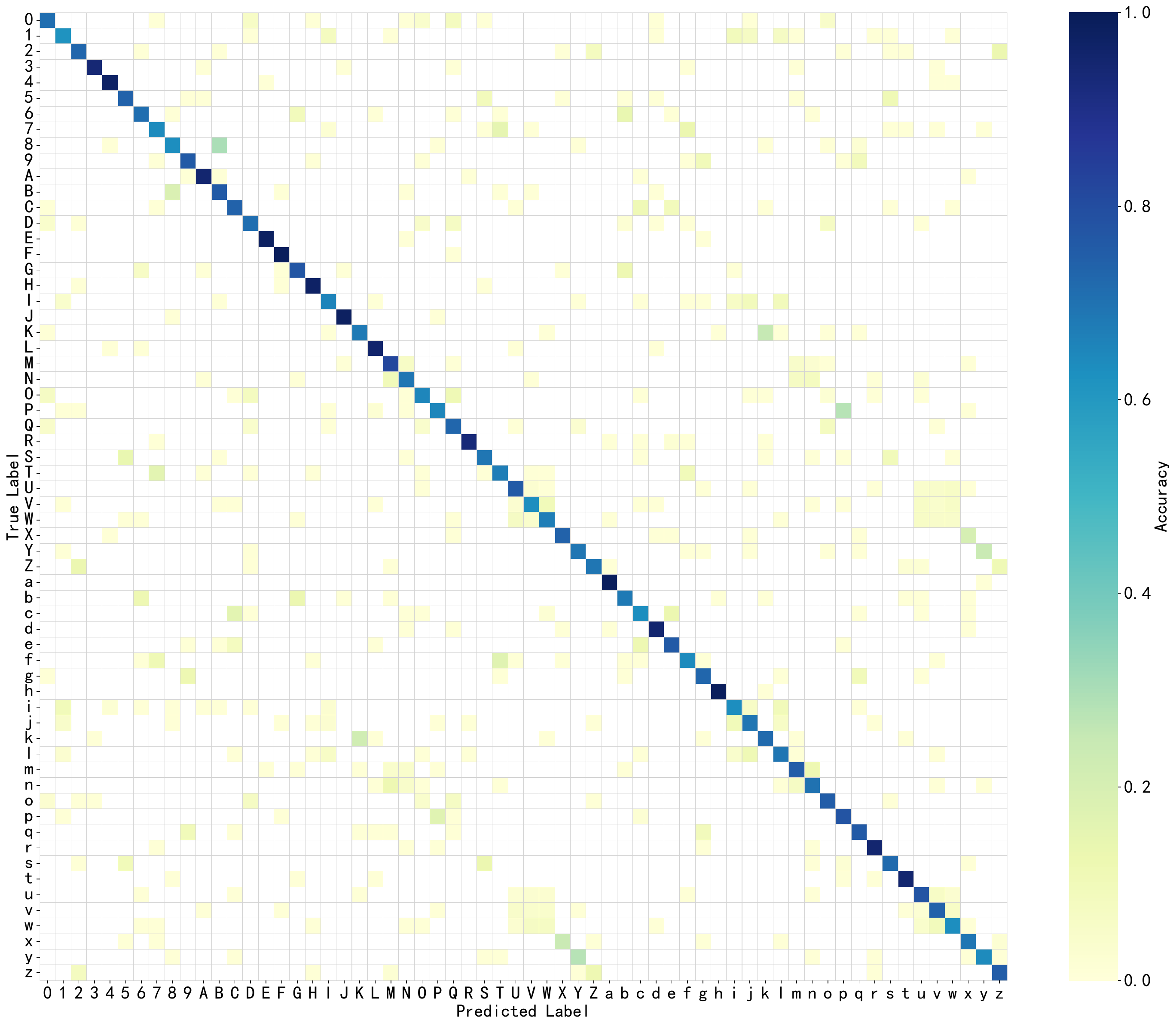}
  \caption{Confusion matrix of character recognition results.}\label{fig: confusion}
\end{figure}

Building on this spatial precision, we assessed the recovery of individual characters. As shown in Figure \ref{fig: handwriting}(a), \Name achieves high-fidelity reconstruction of single-character trajectories. The recovered trajectories accurately preserve the distinct shapes of the original handwriting. We then conducted a quantitative evaluation under two distinct threat scenarios: content recovery (e.g., stealing notes) and biometric trajectory recovery (e.g., signature forgery). In the content recovery scenario, we input 6,200 recovered trajectories (i.e., 100 trajectories per character) into a third-party optical character recognition (OCR) tool Tesseract-OCR \cite{TessOverview}. As shown in Figure \ref{fig: confusion}, the resulting confusion matrix across all 62 characters (A–Z, a–z, and 0–9) yielded an average recognition accuracy of 76.77\%. Errors primarily arise from two understandable sources: non-standard writing variations (e.g., '7' vs. 'T', '9' vs. 'q') and visually similar character pairs (e.g., '0' vs. 'O', 'I' vs. 'l'). In the biometric recovery scenario, we measured geometric similarity between original and recovered trajectories using the Jaccard index. Across 100 random sample pairs, the average Jaccard index similarity reached 0.7374, indicating well spatial overlap.

\begin{table}[!t]
\centering
\caption{Evaluation results of handwriting trajectory recovery.}
\renewcommand{\arraystretch}{1}
\resizebox{\linewidth}{!}{
\begin{tabular}{c|c|c}
\Xhline{1.5pt}
\textbf{Recovery Level}   & \textbf{Recognition Accuracy} & \textbf{Jaccard Index}   \\ \Xhline{1.5pt}
Character-level                &  76.77\%            &                     0.7374                                                                                            \\ 
Word-level            & 58.34\% &                     0.6661 \\ 
\Xhline{1.5pt}
\end{tabular}}
\label{table: handwriting}
\end{table}

Finally, we evaluated the more challenging word-level recovery scenario. The results are presented in Figure \ref{fig: handwriting}(b). For word-level recovery, we performed five consecutive attempts using a third-party recognition tool and adopted the top-5 recovery success rate as the evaluation metric to evaluate the recognizability of the recovered trajectories, defined as a successful attack if any of the first five outputs correctly match the target word. As shown in Table \ref{table: handwriting}, \Name achieves 58.34\% top-5 accuracy, demonstrating that the attack remains effective for practical, real-world word-level recognition tasks. We further applied the Jaccard index to measure trajectory similarity, which remained high at 0.6661. This slight performance degradation compared to character-level recovery is likely attributable to cumulative errors in recovering multiple characters sequentially, as well as to character truncation operations introduced during postprocessing to ensure the independence of each character’s reconstructed trajectory within a word.

\subsection{Impact Factors}

To demonstrate the robustness and practicality of \Name, we conduct experiments to evaluate the impact of various real-world factors on its performance. Note that for each experiment, we follow the same sample sizes as described in Section~\ref{sec: recovery}: 12,000 samples for PIN code recovery, 31,200 samples for keyboard input recovery, 300 samples for application category classification, and 6,200 trajectories for content recovery and 100 sample pairs for biometric recovery at character-level with an 80:20 training-to-testing ratio.

\noindent \textbf{Evaluation on probe-device distance and relative direction.}
To evaluate the practical feasibility of \Name under real-world conditions, we first examine the influence of varying distances and relative direction between the victim device and the EM measurement probe on attack accuracy. Our experiments specifically focus on screen-unlocking PIN code recovery performance, covering a distance range from 5 cm to 25 cm and a relative direction range from $0^{\circ}$ to $180^{\circ}$. For each experiment, we collect 9600 samples (960 samples per digit) for training and 2400 samples (240 samples per digit) for testing. These settings represent realistic attack scenarios in which the attacker may be positioned at varying proximities to the target device.

\begin{figure}[!tbp]
  \centering
  \includegraphics[width=0.9\linewidth]{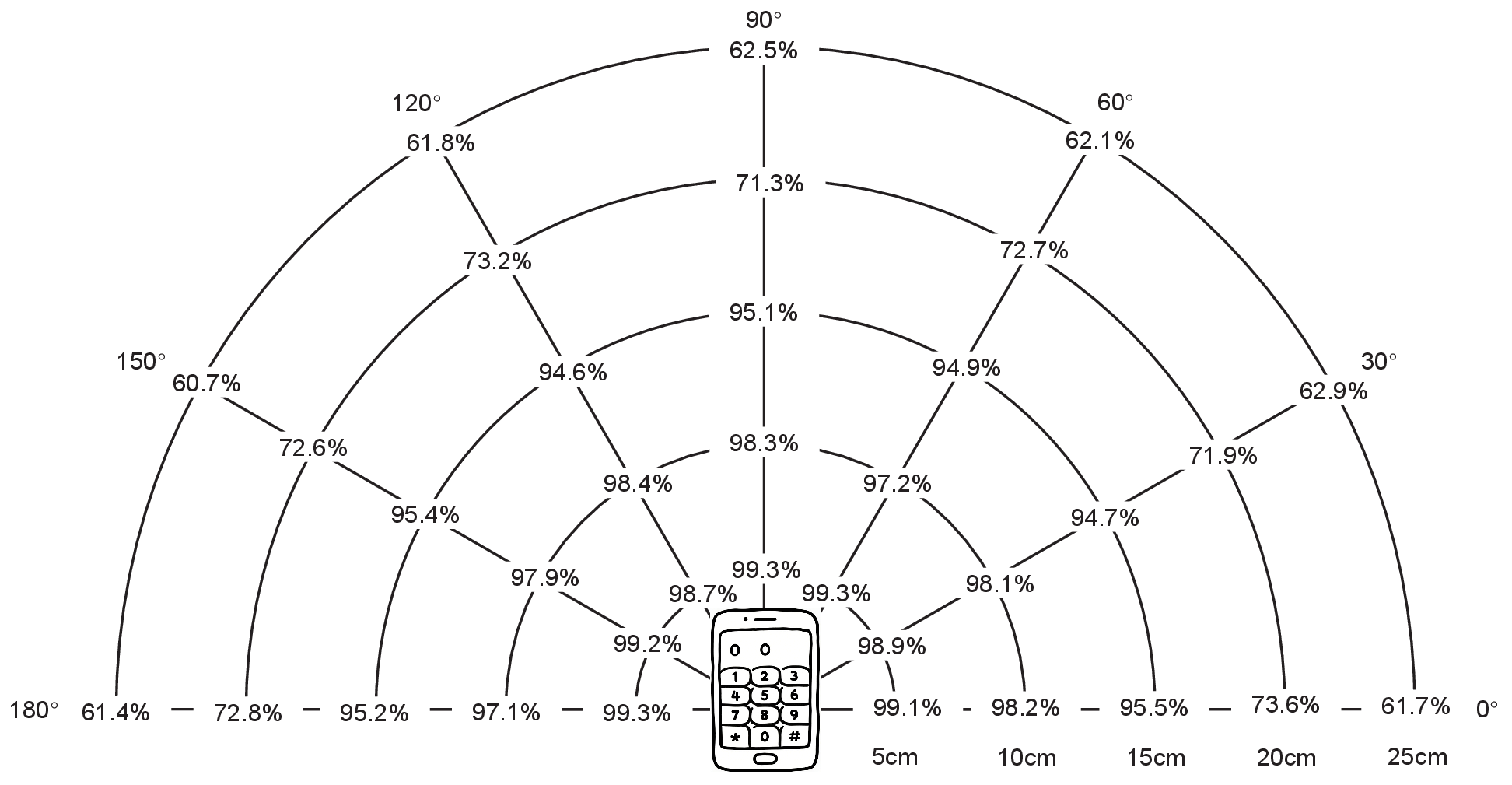}
  \caption{Impacts of probe-device distance and relative direction.}\label{DIstance}
  \vspace{-10pt}
\end{figure}

Fig. \ref{DIstance} illustrates an inverse relationship between attack accuracy and probe-device distance, with classification performance decreasing as the distance increases. This phenomenon can be attributed to the rapid attenuation of electromagnetic field strength with distance, which significantly reduces the signal-to-noise ratio at greater distances. However, \Name still achieves nearly 61\% accuracy for screen-unlocking PIN code recovery at 25cm, which is sufficient for attackers to carry out real-world attacks.

Furthermore, our results indicate that classification accuracy remains nearly consistent across different directions at the same distance. Therefore, the performance of \Name is independent of the relative direction between the probe and the device.

\noindent\textbf{Evaluation on public and private environments.}
Then, we evaluate \Name's attack performance in both public and private environments. As shown in Fig. \ref{environment}, in the public environment, the attacker places a bag containing the probe opposite the victim’s seat at a probe-device distance of 15 cm in a library. In the private environment, the probe is concealed under a wooden table with a thickness of 5 cm, which is positioned in front of the victim in a meeting room.

As shown in Table \ref{tab:envir}, \Name has outstanding attack performance in both public and private environments. Specifically, in the public setting, \Name achieved 90.8\%, 87.5\%, and 84.6\% accuracy for PIN code recognition, keyboard input reconstruction, and application inference, respectively. Simultaneously, it preserves high trajectory recognizability for handwriting, achieving a 72.49\% character recognition accuracy and a Jaccard index of 0.5946. In the private setting, \Name achieved 97.1\%, 96.4\%, and 92.4\% accuracy for the same tasks, alongside a 76.77\% accuracy and a 0.7374 Jaccard index for handwriting. Moreover,  The slightly lower effectiveness of \Name attacks in public environments compared to private environments can be attributed to the higher levels of ambient noise present in public settings. Furthermore, the wooden barrier introduces only minimal performance degradation, as wood products have limited effects on EM signal propagation.

\begin{table}[!tbp]
\caption{Impacts of different environments.}\label{tab:envir}
\renewcommand{\arraystretch}{1.4}
\resizebox{\linewidth}{!}{\begin{tabular}{c|c|c|c|c|c}
\Xhline{1.5pt}
\textbf{Environment} & \textbf{PIN Codes} & \textbf{Keyboard Inputs} & \textbf{Application Categories} & \textbf{Recognition Accuracy} & \textbf{Jaccard Index}\\ \Xhline{1.5pt}
Public               & 90.8\%               & 87.5\%                      & 84.6\%               & 72.49\%               & 0.5946\\
Private              & 97.1\%                & 96.4\%                      & 92.4\%               & 76.77\%               & 0.7374\\ \Xhline{1.5pt}
\end{tabular}}
\end{table}

\noindent\textbf{Evaluation on background applications.}
In addition, given that all background third-party applications are terminated under all the aforementioned experimental conditions, we now evaluate the impact of third-party application interference on \Name. The probe-device distance is fixed at 15 cm and the relative direction is fixed at $90^{\circ}$. During the PIN code entry period, we randomly run up to 10  background applications on the device, including the system’s built-in camera and antivirus application, as well as 8 third-party applications: 2 music applications (e.g., Spotify), 2 video applications (e.g., YouTube), 2 shopping applications (e.g., Amazon) and 2 chat applications (e.g., Wechat).

As Table \ref{tab:background} shows, regardless of the number of third-party applications running in the background, the accuracy of \Name in PIN codes recovery remains consistently around 95\%. The limited impact can be attributed to the fact that these applications only generate noise on the main processor, whereas the EM emanations resulting from touch interaction through the human coupling effect are significantly stronger than those produced by the main processor.

\begin{table}[!tbp]
\caption{Impacts of background applications.}\label{tab:background}
\renewcommand{\arraystretch}{1.4}
\resizebox{\linewidth}{!}{\begin{tabular}{c|c|c|c|c|c|c}
\Xhline{1.5pt}
\textbf{Background Applications} & 0 & 2 & 4 & 6 & 8 & 10 \\ \Xhline{1.5pt}
\textbf{Accuracy} & 95.3\% & 95.3\% & 95.2\% & 95.2\% & 95.2\% & 95.1\%                               \\ \Xhline{1.5pt}
\end{tabular}}
\end{table}

\noindent\textbf{Evaluation on different smartphones.}
Although various smartphones exhibit similar EM leakage security risks, differences in touchscreen parameters may result in distinct characteristics of EM leakage. To evaluate the adaptability of \Name, we follow the same procedures and sample size as before to train and test on three different smartphones, i.e., Xiaomi 10 Pro, Samsung S10, and Huawei Mate 30 Pro.  The probe-device distance is fixed at 10 cm and the relative direction is fixed at $90^{\circ}$ to ensure experimental consistency.

As Table \ref{tab: devices} shows, despite slight variations in accuracy across different phones, the experimental results of our \Name demonstrate that the leakage phenomenon is prevalent across multiple smartphone models. Specifically, on the Huawei Mate 30 Pro, \Name exhibits the most remarkable attack capabilities, where accuracy exceeds 95\% across PIN codes, keyboard inputs, and application categories. Likewise, for continuous handwriting trajectory recovery, the attack consistently achieves a character-level recognition accuracy exceeding 75\% and a Jaccard index greater than 0.68 across all tested manufacturers.

\begin{table}[!tbp]
\caption{{Evaluation results on different smartphones}.}\label{tab: devices}
\renewcommand{\arraystretch}{1.4}
\resizebox{\linewidth}{!}{
\begin{tabular}{c|c|c|c|c|c}
\Xhline{1.5pt}
\textbf{Smartphone} & \textbf{PIN Codes} & \textbf{Keyboard Inputs} & \textbf{Application Categories} & \textbf{Recognition Accuracy} & \textbf{Jaccard Index} \\ \Xhline{1.5pt}
Xiaomi 10 Pro        & 98.1\%               & 95.7\%                   & 93.3\%    &  75.61\%            &                     0.7103                      \\
Samsung S10          & 97.6\%               & 95.0\%                   & 91.7\%     & 74.72\% &                     0.6786                     \\
Huawei Mate 30 Pro   & 99.9\%               & 98.3\%                   & 96.6\%    & 81.34\% &                     0.7425                      \\ \Xhline{1.5pt}
\end{tabular}}
\end{table}

\subsection{Cross-device Testing}
To further demonstrate that \Name poses a practical threat, we launch attacks in a more stringent cross-device scenario, where the attacker trains the classifiers exclusively on their own smartphone and then tests them against the victim's device. We conduct experiments for PIN codes recovery on the four representative smartphones mentioned above (i.e., iPhone X, Xiaomi 10 Pro, Samsung S10, and Huawei Mate30 Pro). For each smartphone model, there are two devices(e.g., two iPhone X phones) used separately for training and testing. On the attacker side, we collect 960 samples per digit from each training device. On the victim side, we collect 240 samples per digit from each testing device. The probe-device distance is fixed at 10 cm and the relative direction is fixed at $90^{\circ}$.

\begin{table}[!tbp]
\caption{{Cross-device results on different smartphones}.}\label{tab: cross}
\renewcommand{\arraystretch}{1.4}
\resizebox{\linewidth}{!}{
\begin{tabular}{c|c|c|c|c}
\Xhline{1.5pt}
\textbf{Smartphone} & \textbf{iPhone X} & \textbf{Xiaomi 10 Pro} & \textbf{Samsung S10} & \textbf{Huawei Mate 30 Pro}\\ \Xhline{1.5pt}
\textbf{Accuracy}   & 96.2\%           & 94.1\%        & 93.3\%   & 97.5\%                       \\ \Xhline{1.5pt}
\end{tabular}}
\end{table}

\begin{figure}[!tbp]
  \centering
  \includegraphics[width=0.9\linewidth]{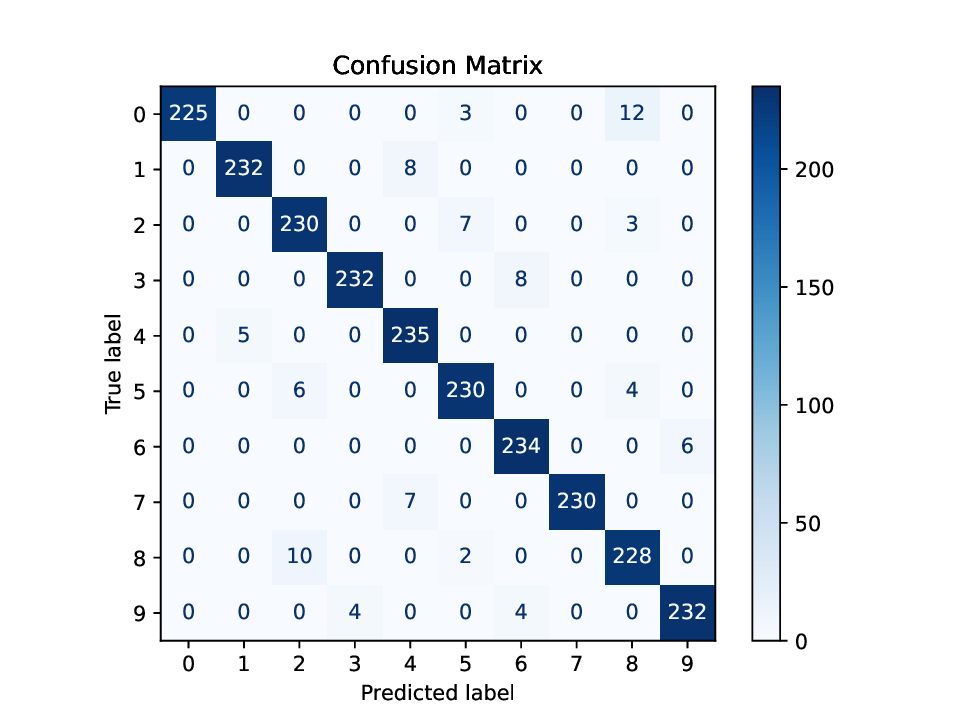}
  \caption{iPhone X Cross-device evaluation on screen-unlocking PIN codes recovery.}\label{cross}
\end{figure}

As shown in Table \ref{tab: cross}, \Name demonstrates strong transferability in the cross-device scenario. The classification accuracy across various smartphone models exhibits only a limited decrease without any model fine-tuning. Specifically, as illustrated in Fig. \ref{cross}, misclassifications within the same numeric column occur more frequently in the cross-device scenario. This further supports our findings in Section \ref{sec: touch location}, which indicate that vertical position variations result in smaller EM signature differences compared to horizontal position variations.

\begin{table*}[t]
\caption{End-to-end attack results.}
\centering \label{tab: end-to-end}
\begin{tabular}{c|c|c|c|c|c}
\Xhline{1.5pt}
\textbf{Scenarios} & \textbf{Smartphone}                & \textbf{Sensetive Information Type} & \textbf{T-1 Results} & \textbf{T-3 Results} & \textbf{T-5 Results} \\ \Xhline{1.5pt}
1                  & \multirow{6}{*}{iPnone X}          & 4-digit PIN Codes                   & 99.00\%              & 100\%                & 100\%                \\
2                  &                                    & 6-digit PIN Codes                   & 96.00\%              & 99.33\%              & 100\%                \\
3                  &                                    & 4-digit Keyboard Inputs             & 94.67\%              & 99.33\%              & 100\%                \\
4                  &                                    & 6-digit Keyboard Inputs             & 94.00\%              & 98.67\%              & 100\%                \\
5                  &                                    & 8-digit Keyboard Inputs             & 85.33\%              & 98.00\%              & 100\%                \\
6                  &                                    & 10-digit Keyboard Inputs            & 84.33\%              & 97.67\%              & 99.33\%              \\ \hline
7                  & \multirow{6}{*}{Xiaomi 10 Pro}     & 4-digit PIN Codes                   & 90.67\%              & 98.67\%              & 100\%                \\
8                  &                                    & 6-digit PIN Codes                   & 84.33\%              & 95.67\%              & 100\%                \\
9                  &                                    & 4-digit Keyboard Inputs             & 86.33\%              & 96.67\%              & 100\%                \\
10                 &                                    & 6-digit Keyboard Inputs             & 77.00\%              & 88.00\%              & 98.33\%              \\
11                 &                                    & 8-digit Keyboard Inputs             & 69.33\%              & 92.67\%              & 95.33\%              \\
12                 &                                    & 10-digit Keyboard Inputs            & 63.67\%              & 85.33\%              & 95.67\%              \\ \hline
13                 & \multirow{6}{*}{Samsung S10}       & 4-digit PIN Codes                   & 85.00\%              & 96.33\%              & 100\%                \\
14                 &                                    & 6-digit PIN Codes                   & 77.67\%              & 94.33\%              & 100\%                \\
15                 &                                    & 4-digit Keyboard Inputs             & 87.33\%              & 96.33\%              & 100\%                \\
16                 &                                    & 6-digit Keyboard Inputs             & 81.00\%              & 95.00\%              & 98.00\%              \\
17                 &                                    & 8-digit Keyboard Inputs             & 72.67\%              & 93.33\%              & 98.67\%              \\
18                 &                                    & 10-digit Keyboard Inputs            & 72.33\%              & 90.33\%              & 98.67\%              \\ \hline
19                 & \multirow{6}{*}{Huawei Mate30 Pro} & 4-digit PIN Codes                   & 99.33\%              & 100\%                & 100\%                \\
20                 &                                    & 6-digit PIN Codes                   & 99.00\%              & 100\%                & 100\%                \\
21                 &                                    & 4-digit Keyboard Inputs             & 95.67\%              & 99.33\%              & 100\%                \\
22                 &                                    & 6-digit Keyboard Inputs             & 94.33\%              & 99.67\%              & 100\%                \\
23                 &                                    & 8-digit Keyboard Inputs             & 96.67\%              & 98.33\%              & 100\%                \\
24                 &                                    & 10-digit Keyboard Inputs            & 90.00\%              & 98.33\%              & 100\%                \\ \Xhline{1.5pt}
\end{tabular}\label{tab: end-to-end}
\end{table*}

\section{End-to-End Attacks}
\label{sec: endtoend}
Although Section \ref{sec: evaluation} has demonstrated \Name's capability to recover individual elements of sensitive information (e.g., single PIN digits, isolated keystrokes, and distinct handwritten characters) as well as continuous word-level handwriting trajectories, it is important to note that for discrete inputs, these component-level results do not directly translate into operational attack effectiveness in real-world scenarios. The practical attack success rate cannot be accurately estimated by simply aggregating individual component accuracies multiplicatively, as actual user interactions involve sequential input patterns and contextual dependencies. Therefore, while the end-to-end viability of handwriting has been established via the word-level trajectory reconstruction in Section VI-D, this section focuses specifically on evaluating the recovery of complete, multi-step discrete interaction sequences.

We conduct a thorough evaluation of end-to-end attack scenarios across four representative smartphone models from leading manufacturers (iPhone X, Xiaomi 10 Pro, Samsung S10, and Huawei Mate30 Pro). For each device, we execute six distinct attack scenarios (a total of 24 scenarios) aimed at reconstructing complete user interaction sequences. These scenarios encompass multiple security-critical operations, beginning with device unlocking using randomly generated 4-digit and 6-digit PIN codes, followed by sensitive data entry in four distinct application categories: (1) app store applications requiring 4-digit inputs, e.g., security verification codes, (2) mobile payment applications involving 6-digit transactions, e.g., payment passwords, (3) shopping applications for 8-digit entries, e.g., coupon redemption codes, and (4) music applications demanding 10-digit inputs, e.g., music search queries. In all scenarios, the probe is hidden directly beneath a wooden table at a probe-device distance of 5 cm, which aligns with our assumption of a private environment. Two devices of the same model are used for training and testing, respectively, and the deep learning model is trained on 10,000 samples.

We implemented an evaluation scheme in which each attack scenario was tested through 300 experimental runs, with each run consisting of 5 consecutive attempts. To evaluate the attack effectiveness, we employed three metrics: (1) top-1 (T-1) success rate, which represents the probability of correctly recovering the sensitive information on the first attempt; (2) top-3 (T-3) success rate, where any successful recovery within the first three consecutive attempts is counted as a successful attack; and (3) top-5 (T-5) success rate, where any successful recovery within the first five consecutive attempts is counted as a successful attack.


Table \ref{tab: end-to-end} provides a comprehensive performance evaluation of our end-to-end attacks across multiple test scenarios. The results indicate exceptional effectiveness, with our method achieving an overall success rate of 99.33\% in reconstructing complete user input sequences across all 24 attack trials when up to five attempts are allowed. Notably, the attack exhibits remarkable consistency, successfully reconstructing all PIN codes across all four tested devices and achieving perfect performance for all sensitive information categories on the Huawei Mate30 Pro within five attempts. Even under the most stringent evaluation condition, where attackers are limited to a single attempt, \Name maintains a substantial success rate of at least 63.67\%, demonstrating its practical viability in real-world scenarios where multiple attempts may be infeasible.

\section{Discussions}
\begin{table}[!tbp]
\caption{{Evalution results of shuffling the keyboard layout on different smartphones}.}\label{tab: countermeasure}
\renewcommand{\arraystretch}{1.4}
\resizebox{\linewidth}{!}{
\begin{tabular}{c|c|c|c|c}
\Xhline{1.5pt}
\textbf{Smartphone} & \textbf{iPhone X} & \textbf{Xiaomi 10 Pro} & \textbf{Samsung S10} & \textbf{Huawei Mate 30 Pro}\\ \Xhline{1.5pt}
\textbf{Accuracy}   & 9.4\%           & 9.7\%        & 9.2\%   & 9.4\%                       \\ \Xhline{1.5pt}
\end{tabular}}
\end{table}

\subsection{Countermeasures}
\textbf{Software-level Countermeasures:} 
Software-level defenses can be deployed via OS updates without hardware modifications. For discrete touch interactions, since the core objective of \Name is to map screen positions to specific semantic inputs, shuffling the virtual keyboard layout disrupts this correlation. As demonstrated in Table \ref{tab: countermeasure}, implementing this randomization causes the accuracy of \Name to plummet to approximately 9\% across all tested devices, which aligns with the expected accuracy of random guessing. 

However, while effective against PINs, such randomization introduces usability challenges and is entirely ineffective against continuous handwriting trajectory recovery, which inherently relies on the absolute physical mapping of finger movements. For continuous inputs, temporal and spatial obfuscation must be applied. The touchscreen driver can dynamically alter the touch sampling rate or insert localized dummy coordinates into the raw touch data pipeline, disrupting the continuous spatiotemporal EM signature without visibly affecting the UI.

\textbf{Hardware-level Countermeasures:}
To fundamentally break the time-to-space mapping ($\Delta t(x)$) exploited by \Name for both discrete and continuous inputs, hardware manufacturers must intervene at the Analog Front End (AFE) level. The most effective approach is scanning sequence randomization. Instead of a fixed, sequential excitation of TX electrodes ($1 \to N_{TX}$), the AFE can employ a pseudo-random shuffling of the activation order in each frame. This hardware-level randomization completely destroys the predictable temporal shifts, rendering the EM emanations of both single taps and continuous writing strokes structurally chaotic to the attacker's model.

Furthermore, differential EM shielding can be enhanced. By integrating thin-film conductive layers or mesh-grid structures specifically tuned to suppress the touchscreen's fundamental scanning frequency, the effective signal-to-noise ratio (SNR) can be minimized at the source, significantly restricting the practical attack distance.

\textbf{Trade-offs and Implementation Cost:}
It is worth noting that while hardware-level defenses offer the highest security, they often involve non-trivial manufacturing costs and potential impacts on touch sensitivity or power consumption. A hybrid approach, combining lightweight UI randomization with optimized AFE scanning logic, represents the most viable path forward for balancing smartphone security, usability, and cost.

\subsection{Limitations and Future Works}
While \Name demonstrates high accuracy and practical feasibility in typical scenarios, several limitations define its current boundaries and open avenues for future research:

\textbf{Extreme Physical Barriers:} As discussed in our threat model, \Name relies on proximate EM propagation. Its efficacy would naturally degrade if the victim's device is placed on heavily shielded surfaces (e.g., thick solid metal desks) or in extreme industrial environments with overwhelming, unpredictable electromagnetic interference.

\textbf{Cumulative Errors in Long Sequences:} As observed in our end-to-end evaluation, the Top-1 exact match accuracy naturally degrades as the length of the input sequence increases (e.g., typing long sentences or complete emails). This is an inherent limitation caused by the accumulation of minor classification errors at individual keystrokes. However, human typing is strictly governed by semantic and syntactic rules. A promising future direction to mitigate this is integrating Large Language Models (LLMs) into the attacker's post-processing pipeline. By feeding the noisy Top-1 sequence or a matrix of Top-k candidate characters into an LLM, the attacker can leverage advanced contextual understanding to automatically autocorrect isolated character errors, effectively bridging the semantic gap and significantly boosting the final recovery rate of meaningful text.

\section{Conclusion}
In this paper, we propose \Name, a practical and completely contactless electromagnetic side-channel attack framework targeting smartphone touchscreens. Breaking away from prior research that requires physical contact or focuses on isolated hardware components, \Name successfully exploits the fundamental scanning mechanism of modern touchscreens under a unified leakage model. Through rigorous spatiotemporal signal alignment and tailored deep learning architectures, our framework can accurately recover both discrete inputs (e.g., PINs and keystrokes) and continuous handwriting trajectories. Comprehensive evaluations in real-world environments demonstrate its high accuracy, cross-unit transferability, and stealthiness. Finally, we provide in-depth hardware and software countermeasures, urging manufacturers to implement scanning sequence randomization to fundamentally mitigate this severe vulnerability.

\section*{Acknowledgment}
During the preparation of this work the authors used Gemini in order to improve the readability and language quality. The tool was not used to generate any scientific content or citations. The authors take full responsibility for the content of the published article.

\bibliographystyle{IEEEtran}
\bibliography{IEEEabrv,sample-base}

\vfill

\end{document}